\documentclass{article}

\usepackage{amsmath}
\usepackage{amsfonts}
\usepackage{pgfplots}
\usepackage{amssymb}
\usepackage[utf8]{inputenc}
\usepackage{tikz}
\usetikzlibrary{positioning}
\usetikzlibrary{matrix,arrows,decorations.pathmorphing}
\usetikzlibrary{patterns}
\usetikzlibrary{through,calc,3d}
\usetikzlibrary{plotmarks}
\usetikzlibrary{pgfplots.groupplots}
\usepackage{varioref}

\usepackage{microtype}\usepackage{mathtools}

\usepackage{subfigure}                  

\usepgfplotslibrary{external}
\newtheorem{definition}{\bf Definition}[section]
\newtheorem{theorem}{\bf Theorem}[section]
\newtheorem{lemma}{\bf Lemma}[section]

\begin{document}

\title{A minimization principle for the description of time-dependent modes associated with
transient instabilities }

\author{Hessam Babaee$^{1,2}$, Themistoklis P. Sapsis$^1$
\thanks{Corresponding author: {sapsis@mit.edu},
Tel: (617) 324-7508, Fax: (617) 253-8689%
 }\\
 $^{1}$Department of Mechanical Engineering, MIT\\
$^{2}$Sea Grant College Program, MIT
}
\date{}



\maketitle
\begin{abstract}
We introduce a minimization formulation for the determination of a finite-dimensional, time-dependent, orthonormal basis that captures directions of the phase space associated with transient instabilities. While these instabilities have finite lifetime they can play a crucial role by either altering the system dynamics through the activation of other instabilities, or by creating sudden nonlinear energy transfers that lead to extreme responses. However, their essentially transient character makes their description a particularly challenging task. We develop a minimization framework that focuses on the optimal approximation of the system dynamics in the neighborhood of the system state. This minimization formulation results in differential equations that evolve a time-dependent basis so that it optimally approximates the most unstable directions. We demonstrate the capability of the method for two families of problems: i) linear systems including the advection-diffusion operator in a strongly non-normal regime as well as the Orr-Sommerfeld/Squire operator, and ii) nonlinear problems including a  low-dimensional system with transient instabilities  and the vertical jet in crossflow. We demonstrate that the time-dependent subspace captures the strongly transient non-normal energy growth (in the short time regime), while for longer times the modes capture the expected asymptotic behavior. 

\end{abstract}


\section{Introduction}

A broad range of complex systems in nature and technology are characterized by the presence of strongly transient dynamical features associated with finite-time instabilities. Examples include turbulent flows in engineering systems
(e.g. Kolmogorov 
\cite{chandler13} and  unstable plane Couette flow
\cite{hamilton95}, reactive flows in combustion  \cite{smooke86, pope97}), turbulent flows in geophysical systems (e.g. climate
dynamics \cite{majda11, majda2000}, cloud process in tropical atmospheric
convection \cite{grab99, grab01}), nonlinear waves (e.g. in optics \cite{akhm13,
arec11} or water waves \cite{onorato13, muller, cousinsSapsis2015_PRE,cousinsSapsis2015_JFM}), and mechanical systems \cite{mohamad2015,arnold_l,vakakis_kourdis,mohamad2015a,Haller2010}.

These systems are characterized by very high dimensional attractors, intense nonlinear energy transfers between modes and broad spectra. Despite their complexity, the  transient features of these dynamical systems are often associated with low-dimensional structures, i.e. a small number of modes,
whose strongly time-dependent character, however, makes it particularly challenging to describe with the classical notion of time-independent modes. This is because these modes, despite their connection with intense energy transfers and transient dynamics, often have low energy and hence they are “buried” in the complex background of modes that are not associated with intense growth or decay but only with important energy. These transient modes often act as “triggers” or “precursors” of higher energy phenomena or instabilities  and a thorough analysis of their properties can have important impact  for i) the understanding of the system dynamics and in particular the mechanisms associated with transient features (see e.g. \cite{schmid2007nonmodal,Tantet2015,Susuki2014}), ii) the prediction  and quantification of upcoming instabilities that are triggered through these low-energy dynamical processes (see e.g. \cite{Susuki2012, cousins_sapsis,cousinsSapsis2015_JFM}),  iii)  the control and suppression of these instabilities by suitably focusing the control efforts in the low energy regime of these transient phenomena (see e.g. \cite{Pastoor2008,tadmor11,Cornelius2013}).

Transient dynamics is central in understanding a wide range of fluid mechanics problems. In the context of hydrodynamic stability, non-normality of the linearized Navier-Stokes operator can cause significant \emph{transient energy growth} \cite{Boberg-and-Brosa,Farrel_92}. It has been well-established now  that the eigenvalue analysis fails to predict the short-time evolution  of  perturbations for convective flows \cite{Trefethen30071993}. Instead,
the transient energy growth can be better understood by analyzing  the pseudospectra of the linearized operator \cite{Tref_Pseudo,doi:10.1137/0153002}. Transition from laminar flow to turbulence, is another
active area of research in fluid mechanics, to which, understanding the transient
dynamical features is crucial. For a finite-amplitude disturbance, \emph{bypass transition} has been observed in wall-bounded shear flows \cite{Schmid:2000aa,RevModPhys.72.603},
in which case the non-normal growth of localized disturbances leads to  small turbulent spots, bypassing the secondary instability process \cite{FLM:339375}.  Recent computational and experimental studies also demonstrate the \emph{sudden transition} from laminar to turbulent motion in pipe flows, where turbulence forms from localized patches called \emph{puffs}  \cite{Moxey04052010,Avila08072011}. On the other hand,
intermittent behavior is the hallmark of turbulent fluid flows.
Turbulent 
\emph{chaotic bursts} appearing in spatially and temporally
localized events can dramatically change the dynamics of the system   \cite{Egolf:2000aa}. Prototype systems that mimic these properties were introduced and analyzed  in~\cite{majda_branicki_DCDS,branic_majda,Majda_filter,mohamad2015}.
Transient dynamics have a fundamental role in the intermittent  behavior of passive tracers as well, where even elementary models without positive Lyapunov exponents \cite{Bourlioux2002,Bourlioux2006} have been able to reproduce intermittent behavior observed in complex models. For such systems, the fundamental role of  the random resonance between Fourier modes of the turbulent velocity field and the passive tracer has been recently illustrated \cite{Tong2015}.

In the context of uncertainty quantification and prediction  a new family of stochastic methods relying on the so-called Dynamical Orthogonality condition was recently developed to deal with the strongly transient features of stochastic systems. The Dynamically Orthogonal (DO) Field Equations \cite{SapsisLermusiaux09, SapsisLermusiaux10} and Dynamically Bi-orthogonal (BO) equations \cite{Hou13a} evolve a subspace according to the system stochastic PDE and the current statistical state of the system. Despite their success in resolving low-dimensional stochastic attractors for PDES \cite{sapsis11a,sapsisdijkstra,sapsis_DONS}, these methodologies are often  too expensive to implement for high dimensional systems (e.g. DO or BO require the simultaneous solution of many PDEs). In addition, in order to obtain an accurate description of the time-dependent dynamics many modes should be included in the analysis and the computational cost increases very rapidly (especially for systems with high complexity). 

Our aim in this work is to develop a method that will generate adaptively a time-dependent basis that will capture strongly transient phenomena. This approach will rely on system observables obtained either through high fidelity numerical solvers or measurements, as well as the linearized equations of the system. The core of our approach is a minimization principle that will seek to minimize the distance between the local vector field of the system, constrained over the direction of the time-dependent modes, and the rate of change of the time-dependent modes. A direct minimization of the defined functional will result in evolution equations for the optimally time-dependent (OTD) basis elements. For systems characterized by transient responses these modes will adapt according to the (independently) computed or measured  system  history in a continuous way, capturing at each time the transiently most unstable directions of the system.
For sufficiently long times where the system reaches an equilibrium we prove that the developed equations provide the most unstable directions of the system in the asymptotic limit.

We demonstrate the developed approach over a series of applications, including linear and nonlinear systems. As a first example we consider the advection-diffusion operator where we show how the OTD basis captures the directions associated with the non-normal behavior. The second example involves  the Orr-Sommerfeld/Squire operator that governs
the evolution of infinitesimal disturbances in parallel viscous flows. Our goal here is also the computation of time-dependent modes that explain the transient growth of energy due to non-normal dynamics. The third problem involves low-dimensional dynamical  system as well as the  nonlinear transient dynamics of a jet in cross flow. Using the developed framework we compute the modes associated with the transient but also asymptotically unstable directions of the phase space and we assess their time-dependent stability properties. 

\section{Optimally time-dependent (OTD) basis for transient instabilities}

Let the dynamical system
\begin{align*}
\dot{z} & =F\left(z,t\right),
\end{align*}
defined on a state space $A\subset\mathbb{R}{}^{n}.$ We denote by
$S_{t}\left (z_{0}\right)$ the position of the trajectory at time $t$ 
that is initiated on $z_{0}$. Also, let 
\begin{equation}\label{linearized_dynamics}
\dot u=L({S_{t},t)}u,\qquad \text{with} \quad L(z,t)= \nabla
_{z}F(z,t),
\end{equation}
denote the linearized dynamical system around the trajectory $S_{t}$ and let the inner product between two elements $z_1$ and $z_2$ be denoted as $z_1  \cdot z_2$. The linear time-dependent dynamical 
system represented by equation (\ref{linearized_dynamics}) has the solution 
\begin{equation}
u(t) = \Phi_{t_0}^t u(t_0),
\end{equation}
where $\Phi_{t_0}^t$ is the propagator that maps the state of the system at time $t_0$ to $t$. The propagator can be represented as the ordered product of infinitesimal propagators 
\begin{equation}
\Phi_{t_0}^t = \lim_{\delta t \to 0} \prod_{j=1}^n{e^{L(S_{t_j},t_j)\delta t}},
\end{equation}
where $t_j$ lies in $t_0+(j-1)\delta t < t_j < t_0 +j\delta t$ with $t=t_0+n\delta t$.
Our aim  is to evolve a  basis $u_{i}, i=1,...,r$, i.e. a set of
time-dependent, orthonormal modes, so that  $u_i(t)$ optimally follows $\Phi_{t_0}^t u_i(t_0)$  for all times.  To achieve this goal we formulate
the following quantity, which measures
the distance between the action of infinitesimal propagator $e^{L(S_{t_j},t_j)\delta
t}$ on  an orthonormal basis $u_i(t)$ and  $u_i(t+\delta t)$. We have:

\begin{align}\label{eq:functional_0}
 \mathcal{F} =\frac{1}{\left(\delta t\right)^2} \sum_{i=1}^{r} \big\|u_i(t+\delta t)- \Phi_{t}^{t+\delta t}u_i(t)\big\|^2, \quad \quad \delta t \rightarrow 0,
 \end{align}
where $U(t)=[u_{1}\left(t\right),u_{2}\left(t\right),...,u_{r}\left(t\right)]$
is an arbitrary and time-dependent orthonormal basis, i.e. for every time
instant
it satisfies the orthonormality condition:
\begin{align}
u_{i}\left(t\right)\cdot u_{j}\left(t\right) & =\delta_{ij},\,\, i,j=1,...,r.\label{eq:orthono}
\end{align} 
We observe that in the  functional given by equation (\ref{eq:functional_0}):
\begin{align*}
u_i(t+\delta t) &= u_i(t)+\delta t \dot{u}_i + \mathcal{O}(\delta t^2),\\
\Phi_{t}^{t+\delta t} &= e^{L(S_t,t)\delta t} = I+\delta t L(S_t,t) + \mathcal{O}(\delta t^2),
\end{align*}
where $I$ is the identity matrix.
Replacing the above equations into the functional given by equation (\ref{eq:functional_0}) results in:
\begin{align}
\mathcal{F}(\dot u_1,\dot u_2, \dots,\dot u_r) 
& =\sum_{i=1}^{r}\left\Vert \frac{\partial u_{i}\left(t\right)}{\partial
t}-L({S_{t},t)u_i(t)}\right\Vert ^{2},\label{eq:functional}
\end{align}

\begin{figure}
\centering
\includegraphics[width=0.35\textwidth]{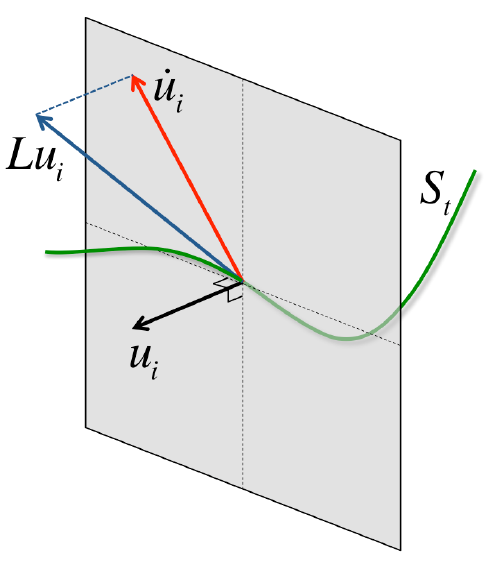}
\caption{Illustration of the distance function $\mathcal{F}$ utilized to define the time-dependent modes.}
\label{fig1}
\end{figure}

In Figure \ref{fig1} we illustrate the distance function $\mathcal{F}$. For each direction $u_i$ we have, due to the normalization property $\left\Vert u_i \right\Vert=1$, the rate of change $\dot u_i$ lying in the orthogonal complement of $u_i$. Under this constraint we choose as $\dot u_i$ the vector that is closest to the image of $u_i$ under the effect of the operate $L$. 

We emphasize that the minimization of the function (\ref{eq:functional}) is considered only with respect to the time-derivative (rate of change) of the basis, $\dot U(t),$ instead of the basis $U(t)$ itself. This is because we do \textit{not} want to optimize the subspace that the operator is acting on, but rather find an optimal set of vectors, $\dot U(t),$ that best approximates the  linearized dynamics in the subspace $U$. We then solve the resulted equations and compute $U(t)$. We will refer to these modes as the \textit{optimally time-dependent} (OTD)
modes, and the space that these modes span as the OTD subspace.

Note that a direct minimization of the function (\ref{eq:functional}), over finite-time intervals, with respect to the modes $U(t),$ would result the well-known Euler-Lagrange equations. However, in this case the emphasis is put on finding the optimal `input' to the operator $L$ in order for the function (\ref{eq:functional}) to be minimized, while here our aim is different, i.e. we seek to find an optimal set of vectors that approximate $L$ in the best possible way, when $L$ is acting on $U$. 

To summarize the above discussion, we are considering the following minimization problem: 
 \begin{equation*}
 \mbox{For a trajectory} \ \ \ S_{t} \ \  \mbox{ find} \quad \dot U(t) \quad \mbox{such that} \quad \mathcal{F} \left(\dot U \right) \quad \mbox{is minimized.}
 \end{equation*} 
 

In the reminder of this paper we will demonstrate that this minimization principle allows to capture any arbitrary and transient kind of growth caused
by the linear operator, e.g. non-normal or exponential.
In what follows we obtain a differential formulation for the evolution equations
that correspond to the proposed minimization problem.

\subsection{Evolution equations for the time-dependent modes}

Before we proceed to the minimization of the function given by equation (\ref{eq:functional}),
we express the orthonormality constraints for the basis elements in terms
of their time derivatives. By differentiating over time, we have
\begin{align}
\frac{\partial u_{i}\left(t\right)}{\partial t}\cdot u_{j}\left(t\right)+\frac{\partial
u_{j}\left(t\right)}{\partial t}\cdot u_{i}\left(t\right) & =0,\,\, i,j=1,...,r.\label{eq:orthon_dyna}
\end{align}
The above condition is satisfied if the following condition is valid:
 \begin{align}
\frac{\partial u_{i}\left(t\right)}{\partial t}\cdot u_{j}\left(t\right)
& =\varphi_{ij}\left(t\right),\label{eq:angle_cond}
\end{align}
where $\varphi_{ij}(t)$ is any arbitrary function for which $\varphi_{ij}=-\varphi_{ji}$.
Clearly, for any choice of $\varphi_{ij}(t)$
the above condition guarantees that if $\left\{ u_{i}\left(t_{0}\right)\right\}
_{i=1}^{r}$
is initially orthonormal, it will remain orthonormal for all times.
As we will see the choice of $\varphi_{ij}$ does not change the evolved subspace.
However, it allows for different formulations of the evolution equations.
Using the constraint (\ref{eq:angle_cond}) we have the following Theorem.
\begin{theorem} 
  The minimization principle (\ref{eq:functional})
defined within the basis elements that satisfy the constraint (\ref{eq:angle_cond})
is equivalent with the set of evolution equations
\begin{align}
\frac{\partial u_{k}\left(t\right)}{\partial t} & =L({S_{t},t)u_k(t)}-\sum_{j=1}^{r}\left(L({S_{t},t)u_k(t)}\cdot
u_{j}\left(t\right)-\varphi_{kj}(t)\right)u_{j}\left(t\right),\,\, k=1,...,r,\label{eq:DO_basis}
\end{align}
where $\varphi_{ij}$ is an arbitrary function for which $\varphi_{ij}=-\varphi_{ji}$.

\end{theorem}
\textbf{Proof:} We first formulate the minimization problem
that also takes into account the appropriate number of Lagrange multipliers, $\lambda_{ij}\left(t\right),$with $i,j=1,...,r$.
In this way we obtain:
\begin{align}
\mathcal{G_{\varphi}}\left(\dot U\left(t\right);L({S_{t},t)}\right)
& =\sum_{i=1}^{r}\left(\frac{\partial u_{i}\left(t\right)}{\partial
t}-L({S_{t},t)u_i(t)}\right)^{2}+\sum_{j=1}^{r}\lambda_{ij}\left(t\right)\left(\frac{\partial
u_{i}\left(t\right)}{\partial t}\cdot u_{j}\left(t\right)-\varphi_{ij}\left(t\right)\right).\label{eq:variati_princ_mod}
\end{align}
We consider the first derivative with respect to each
$\dot{u}_{i}\left(t\right)$ to obtain the set of equations
\begin{align*}
\frac{\partial\mathcal{G_{\varphi}}}{\partial\dot{u}_{k}} & =2\left(\frac{\partial
u_{k}\left(t\right)}{\partial t}-L({S_{t},t)u_k(t)}\right)+\sum_{j=1}^{r}\lambda_{kj}\left(t\right)u_{j}\left(t\right).
\end{align*}
To obtain an extremum we need the right hand side of the last equation to vanish:
\begin{align}
\frac{\partial u_{k}\left(t\right)}{\partial t} & =L({S_{t},t)u_k(t)}-\frac{1}{2}\sum_{j=1}^{r}\lambda_{kj}\left(t\right)u_{j}\left(t\right),\label{eq:inter_equ_mod}
\end{align}
which should be solved together with  condition (\ref{eq:angle_cond}).
We take the inner product of equation (\ref{eq:inter_equ_mod}) with
mode $u_{l}\left(t\right)$ and obtain
\begin{align*}
\frac{\partial u_{k}\left(t\right)}{\partial t}\cdot u_{l}\left(t\right)
& =L({S_{t},t)u_k(t)}\cdot u_{l}\left(t\right)-\frac{\lambda_{kl}\left(t\right)}{2}=\varphi_{kl}\left(t\right).
\end{align*}
Using the last equation and substituting  $\lambda_{kl}(t)$ in (\ref{eq:inter_equ_mod})
will result in the evolution equations (\ref{eq:DO_basis}). This completes the proof.

In a more compact form the evolution equation for a finite-dimensional operator
 $L \in \mathbb{R}^{n\times n }$ can be obtained, where we  express the OTD subspace in a matrix $U \in \mathbb{R}^{n\times
r}$ whose $i^{\mbox{th}}$ column is $u_i$. The function $\varphi_{ij}$ is correspondingly expressed in the matrix notation as $\varPhi \in \mathbb{R}^{r \times r}$ with $\varPhi=\{\varphi_{ij}\}_{i,j=1}^r$. Thus, the evolution equation for the OTD modes can be expressed as: 
\begin{equation}\label{dynamical_sys_sub_withphi}
\frac{\partial U}{\partial t} = LU - U(U^T
L U-\varPhi),
\end{equation}
where $(\ )^T$  denotes  the transpose of a matrix.

Now we define the \emph{reduced operator}
 $L_r(t) \in \mathbb{R}^{r\times r}$ that is obtained by projecting the original
operator onto the  subspace  $U(t)$. Thus,
\begin{equation}\label{reduction}
L_r = U^T LU.
\end{equation}
Therefore the OTD equation could be equivalently expressed as:
\begin{equation}\label{dynamical_sys_sub_withLr}
\frac{\partial U}{\partial t} = LU - U(L_r-\varPhi).
\end{equation}
In what follows we will need to the definition of equivalence between two subspaces.
 \begin{definition}
The two OTD subspaces $U  \in \mathbb{R}^{n\times
r}$ and  $W  \in \mathbb{R}^{n\times
r}$  are equivalent at time $t$ if there exists a transformation matrix $R 
\in \mathbb{R}^{r
\times r}$ such that $U(t)=W(t)R$, where $R$  is an orthogonal rotation matrix, i.e. $R^TR=I$.
\end{definition}
In the following we show that the evolution of two OTD subspaces  $U(t)  \in \mathbb{R}^{n
\times r}$ and  $W(t)  \in \mathbb{R}^{n
\times r}$ under different choice of $\varPhi(t)$, which are  initially equivalent (at $t=0$),    will remain equivalent  for every time $t>0$, \emph{i.e.} $U(t)=W(t)R(t),$  where  $R(t)  \in \mathbb{R}^{r
\times r}$ is an orthogonal rotation matrix    governed by the  matrix differential equation:
  \begin{align}\label{eq:orthogonal_rot}
\frac{dR}{dt}&=R \varPhi_U-\varPhi_WR,\\ \nonumber
R(0)&=R_0,
\end{align}
with $ \varPhi_U \in \mathbb{R}^{r
\times r} $ and $\varPhi_W \in \mathbb{R}^{r
\times r}$ being the two different choices of $\varPhi$ for the evolution of $U$ and $W$, respectively, while $R_0$ in the initial orthogonal rotation matrix, \emph{i.e.} $U(0)=V(0)R_0$.
We first prove the following Lemma.
\begin{lemma} 
The solution $R(t)$ to the matrix differential equation given by equation (\ref{eq:orthogonal_rot}), remains an orthogonal rotation matrix for every time $t>0$ given that the initial condition $R(0)$ is an orthogonal rotation, \emph{i.e.} $R(0)^T R(0)=I$, and $\varPhi_U$ and $\varPhi_W$ are skew-symmetric matrices. 
\end{lemma}
\textbf{Proof:}
We show that $\dfrac{d (R^T R)}{dt}=0$ for every $t>0$ :
\begin{align*}
\frac{d (R^T R)}{dt}&=\dot{R}^T R+R^T\dot{R}\\
                    &=(R \varPhi_U-\varPhi_W R)^T R + R^T(R \varPhi_U-\varPhi_W R)\\
                    &=-\varPhi_U R^TR+R^T\varPhi_W R+R^TR\varPhi_U-R^T\varPhi_WR\\
                    &=R^TR\varPhi_U-\varPhi_U R^TR.
\end{align*}
Clearly $R^T R=I$ is a fixed point for the above equation.

Next we prove that the for a given dynamical system, the OTD subspaces that are initially equivalent, remain equivalent for all times. 
 \begin{theorem}\label{Thm:Equi_Phi} 
 Suppose that  $U(t)  \in \mathbb{R}^{n
\times r}$ and  $W(t)  \in \mathbb{R}^{n
\times r}$  satisfy the evolution equation (\ref{dynamical_sys_sub_withphi}) with different
choices of $\varPhi(t)$ functions denoted by $\varPhi_U(t) \in \mathbb{R}^{r
\times r}$ and $\varPhi_W(t) \in \mathbb{R}^{r
\times r}$ respectively.
We also assume that the two bases are initially equivalent, \emph{i.e.
} $U(0)=W(0)R_0$, where $R_0 \in \mathbb{R}^{r
\times r}$ is an orthogonal rotation matrix. Then the subspaces $U(t)$ and $W(t)$
 are equivalent for $t>0,$ with a rotation matrix $R(t) \in \mathbb{R}^{r
\times r}$ governed by the  matrix differential
equation  (\ref{eq:orthogonal_rot}).

\end{theorem}
\textbf{Proof:}
We plug  $U(t)=W(t)R(t)$ into the OTD equation for $U(t)$:

\begin{equation*}
\dot{W}R+W\dot{R}=LWR-WR(R^TW^TLWR-\varPhi_U).
\end{equation*}
Multiplying both sides from the right with $R^T$ and using the identity $RR^T=I, $  results in:
\begin{equation*}
\dot{W}=LW-W(TW^TLW-R\varPhi_UR^{T}+\dot{R}R^T).
\end{equation*}
Now we substitute $\dot{R}$ from equation (\ref{eq:orthogonal_rot}) into the above equation to obtain:
\begin{equation*}
\dot{W}=LW-W(W^TLW-\varPhi_W),
\end{equation*}
which is the evolution equation for the OTD basis $W(t)$. This completes the proof.


Theorem \ref{Thm:Equi_Phi}  implies that the difference between the two bases will evolve
along directions already contained in the initially common subspace.
To this end, both bases will continue to span the same subspace and
the  variation between the two is only an internal rotation.
Therefore, the two family of equations 
will result in the same time-dependent subspace. There are multiple choices for the function $\varphi_{ij}$ and we now examine a special one.  
\subsubsection{The Dynamically Orthogonal formulation}
The simplest choice for the function $\varphi$ in (\ref{eq:angle_cond})
is $\varphi_{ij}=0$ for all $i,j$. The resulted evolution equations in
this case will have the form
\begin{equation}\label{dynamical_sys_sub}
\frac{\partial U}{\partial t} = QLU ,
\end{equation}
  where  $Q := I - UU^T $
is the orthogonal projection  operator  onto the subspace $U$. Note that $\varphi_{ij}=0$ corresponds to the dynamical orthogonality
(DO)\ condition \cite{SapsisLermusiaux09, sapsis11a} that has been employed
to derive closed equations for the
solution of stochastic PDEs. In this case uncertainty is resolved
only along specific modes that evolve with time by projecting the original
equation of the system over these directions. The evolution of
these modes (stochastic subspace) is  done according to equations
derived using the DO condition and they have the general form of system
(\ref{dynamical_sys_sub}). The equivalence of system (\ref{dynamical_sys_sub})
with the minimization problem (\ref{eq:functional}) provides a clear
interpretation for the evolution of the DO modes.

\subsection{Steady linearized dynamics}

Here we consider the special case where $L$ is a time-independent
operator. We prove that the basis defined through the introduced minimization principle
will asymptotically span the eigenvectors of $L$ \textit{associated with the  most intense
instabilities} (i.e. eigenvalues with largest real part).
In particular we have the following theorem:

\begin{theorem}\label{Thm:SLD}
 Let  $L \in \mathbb{R}^{n\times n }$  be a steady and diagonalizable operator that represents the linearization of an autonomous dynamical system. Then

i) Equation (10) has  $\binom{n}{r} = \frac{n!}{(n-r)! r!}$ equilibrium states  that  consist of all the \(r\)-dimensional subspaces in the span of  \(r \) distinct eigenvectors of \(L\). 

ii) From all the equilibrium states there is only one that is a stable solution for equation (10). This is given by the subspace spanned by the eigenvectors of $L$ associated with the \(r\) eigenvalues having the largest real part.

\end{theorem}
\textbf{Proof:}
 Let $L\Phi = \Phi \Lambda$  where $\Phi$ is the matrix of eigenvectors of the operator $L$ with the column of $\Phi \in \mathbb{R}^{n\times n}$
 being the eigenvectors: $\Phi=\{\phi_1 | \phi_2 | \dots |\phi_n\} $, and $\Lambda  \in \mathbb{R}^{n\times n}$ is the diagonal eigenvalue matrix whose entries are:
  $\Lambda = \mbox{diag}(\lambda_1, \lambda_2, \dots, \lambda_n)$.

\textbf{i)}  First, we show that a subspace $U_0$ that is in the span of
precisely $r$ eigenvectors of the operator $L$, is an  equilibrium state for  equation
 (\ref{dynamical_sys_sub}). Without loss of generality, we consider the first $r$ eigenvectors to span such space, \emph{i.e.} $U_0 \in \Phi_r=\mbox{span} \{\phi_1 | \phi_2 | \dots |\phi_r\}$ associated with $\Lambda_r = \mbox{diag}(\lambda_1, \lambda_2, \dots, \lambda_r) $, and therefore  $U_0$  can be expressed in eigenvector coordinates as:
  $U_0 = \Phi_r \kappa_0$ where $\kappa_0 \in \mathbb{R}^{r\times r}$ is the projection coefficients.
   Therefore $L U_0 = L \Phi_r \kappa_0 = \Phi_r \Lambda_r \kappa_0 \in \mbox{span} \{\phi_1 | \phi_2 | \dots |\phi_r\}$.
   As a result,  $L U_0$ is in the null space of the orthogonal projector $Q$, and thus $QL U_0 = 0$. 
   This result is independent of the choice of eigenvectors. To this end, we note that, for an operator $L$  with distinct eigenvalues, there exists a number of  $\binom{n}{r} = \frac{n!}{(n-r)! r!}$ of such equilibrium states.

\textbf{ii)}
Next, we show that from all the equilibrium states $U_0$, the subspace that is spanned by the eigenvectors associated with the  largest eigenvalues, is the only stable  equilibrium.  First, we  investigate the  stability of $U_0 \in \Phi_r$. We denote the complement of the space $\Phi_r$ with $\Phi_{r^c} = \mbox{span} \{\phi_{r+1} | \phi_{r+2} | \dots |\phi_n\}$ associated with the corresponding eigenvalues of $ \Lambda_{r^c} = \mbox{diag}(\lambda_{r+1}, \lambda_{r+2}, \dots, \lambda_n) $. We consider  a perturbation $U' \in~\mathbb{R}^{n\times r} $  that belongs to the orthogonal complement of  $U_0$, i.e. $U' \in{U_{0}^C}$:
   \begin{equation*}
U(t) = U_0 + \epsilon U'(t), \quad \quad U'(t) \perp U_0.
   \end{equation*}
We note that the orthonormality condition for $U(t)$, \emph{i.e.} $U(t)^T U(t)=I$, is  satisfied for $\epsilon<<1$:
 \begin{align*}
U(t)^T U(t)&= \big(U_0 + \epsilon U'(t)\big )^T\big(U_0 + \epsilon U'(t)\big )\\&= U_0^TU_0 + \epsilon(U_0^TU'(t)+U'(t)^TU_0)+\epsilon^2 U'(t)^TU'(t)\\
&=I + \epsilon^2 U'(t)^TU'(t) \\
&\simeq I, \quad \quad \mbox{for} \ \epsilon<<1.
   \end{align*}
In the above equation, we use the orthonormality condition of $U' \perp U_{0}$ that implies: $U_0^TU'(t)=U'(t)^TU_0=0$. Moreover, since  $U_0^TU'(t)=0,$ we immediately obtain:
\begin{equation}\label{Ortho_req}
 U_0^T \frac{\partial U'(t)}{\partial t}=0 \quad \quad t\geq 0,
 \end{equation}
 which requires the evolution of the perturbation,  \emph{i.e.} $\partial U'(t)/\partial t $, to  remain
orthogonal to $U_0$ for all times.
  Now,  linearizing the evolution equation stated in equation (\ref{dynamical_sys_sub}) around the equilibrium state $U_0$ yields:
   \begin{equation*}
\frac{\partial U'}{\partial t} = Q L U' - U_0  U'^T L U_0
-U' U_0^T L U_0.
   \end{equation*}
   In the above equation, the term $U'^T L U_0= 0$, since $L U_0 \in \mbox{span}
\{\phi_1 | \phi_2 | \dots |\phi_r\}$ and $U' \perp U_0$.
Therefore the evolution equation for the perturbation equation becomes:
  \begin{equation}\label{dynamical_sys_sub_lin}
\frac{\partial U'}{\partial t} = Q L U' 
-U' U_0^T L U_0.
   \end{equation}
Next, we transform the evolution equation (\ref{dynamical_sys_sub_lin}) into the eigenvector coordinates.   The perturbation $U'$ can be expressed in eigenvector coordinates  as:
$U' = \Phi \kappa'$, where $ \kappa' \in  \mathbb{R}^{n\times r }$. Replacing   $U_0 = \Phi_r  \kappa_0$ and $U' = \Phi \kappa'$ into equation (\ref{dynamical_sys_sub_lin}) results in:
 \begin{equation}\label{dynamical_sys_sub_lin_eig}
\Phi\frac{d \kappa'}{d t} = Q \Phi \Lambda \kappa' - \Phi \kappa'  \kappa_0^T \Phi_r^T \Phi_r \Lambda_r \kappa_0.
   \end{equation}
    Having assumed that $L$ is non-deficient, we multiply both sides of  equation (\ref{dynamical_sys_sub_lin_eig}) by $\Phi^{-1}$:
 \begin{align}
\frac{d \kappa'}{d t} &= \Pi \Lambda \kappa' -  \kappa'  \kappa_0^T
\Phi_r^T \Phi_r \Lambda_r \kappa_0\\ \nonumber
 &= \Pi\Lambda \kappa' -  \kappa'  \kappa_0^{-1}  \Lambda_r
\kappa_0,
   \end{align}
where we introduced $ \Pi= \Phi^{-1}Q\Phi$, and used  the  orthonormality
condition of ${U_0^T} U_0 = \kappa_0^T \Phi_r^T  \Phi_r\kappa_0 = I$.
 We then multiply both sides by $\kappa_0^{-1} $ to obtain:
   \begin{equation}\label{pert_aux}
\frac{d \rho'}{d t} = \Pi \Lambda \rho' -  \rho'   \Lambda_r,
   \end{equation}
   where $\rho' =  \kappa' \kappa_0^{-1}$.
   The perturbation $\rho'$  can be  decomposed as:
   \begin{equation}
\rho' = \begin{pmatrix}
 \rho'_r \\
 \rho'_{r_c}
\end{pmatrix},
   \end{equation}
where $\rho'_r \in \mathbb{R}^{r\times r}$ and $\rho'_{r^c} \in \mathbb{R}^{r^c\times
r}$, with $r+r^c = n$. The matrix $\Pi$ can be written as:
 \begin{align}\label{Pi}
 \Pi  &= I_{n\times n} - \Phi^{-1}\Phi_r\kappa_0\kappa_0^T\Phi_r^T\Phi\\ \nonumber
      &= I_{n\times n} - \Phi^{-1}\Phi_r\kappa_0\kappa_0^T\Phi_r^T \begin{pmatrix}
 \Phi_r  & \Phi_{r^c} \\
\end{pmatrix}.
 \end{align} 
In the above  equation, we note that:
 \begin{align}
\Phi^{-1}\Phi_r &=  \begin{pmatrix}
 I_{r\times r}   \\
 0_{r_c\times r}
\end{pmatrix}. \label{Pi_aux1} \\
\kappa_0\kappa_0^T\Phi_r^T \begin{pmatrix}
 \Phi_r  & \Phi_{r^c}
\end{pmatrix} &=  \begin{pmatrix}
 \kappa_0\kappa_0^T\Phi_r^T\Phi_r\kappa_0\kappa_0^{-1}  & \kappa_0\kappa_0^T\Phi_r^T\Phi_{r^c} \\
\end{pmatrix} \label{Pi_aux2}\\ \nonumber
&= \begin{pmatrix}
I_{r\times r}  & \Theta_{r\times r_c}
\\
\end{pmatrix}. 
 \end{align} 
In the last equation, we used the  orthonormality
condition of ${U_0^T} U_0 = \kappa_0^T \Phi_r^T  \Phi_r\kappa_0 = I$  and introduced 
$\Theta = \kappa_0 \ \kappa_0^T \Phi_r^T \Phi_{r^c}$.
Replacing equations (\ref{Pi_aux1}) and (\ref{Pi_aux2})
into equation (\ref{Pi}) results in:
\begin{equation}\label{Pi_Final}
\quad \quad \quad  
\Pi  = \begin{pmatrix}
 0_{r\times r} & \Theta_{r\times r_c}   \\
0_{r_c\times r} & I_{r_c\times r_c}   \\
\end{pmatrix}.
\end{equation}
By replacing equation (\ref{Pi_Final}) into equation (\ref{pert_aux}),  the evolution equation becomes:
\begin{equation}\label{evol_perturb_aux}
   \frac{d }{d t} \begin{pmatrix}
 \rho'_r \\
 \rho'_{r_c}
\end{pmatrix} = \begin{pmatrix}
 0 & \Theta\Lambda_{r^c}   \\
 0 & \Lambda_{r^c}   \\
\end{pmatrix} \begin{pmatrix}
 \rho'_r \\
 \rho'_{r_c}
\end{pmatrix} - \begin{pmatrix}
 \rho'_r \Lambda_{r} \\
 \rho'_{r_c} \Lambda_{r}
\end{pmatrix}.
   \end{equation}
   The evolution equation for the perturbation must satisfy the orthogonality constraint    $\partial U'(t)/\partial t \perp U_0 $ expressed by equation (\ref{Ortho_req}).  The orthogonality constraint
requires that:
\begin{equation*}
U_0^T\frac{\partial U'}{\partial t} = U_0^T( Q L U' 
- U' U_0^T L U_0)=0.
   \end{equation*}
Since $U_0 \in \mathbb{R}^{n\times r}$, the above orthogonality condition, in fact, imposes $r$ constraints on the evolution of perturbation $\rho'(t)$. In the following, we simplify these constraints. In the above equation:
\begin{align*}
 U_0^T( Q L U' 
- U' U_0^T L U_0) &=  \kappa_0^T \Phi_r^T( \Phi \Pi \Phi^{-1} \Phi\Lambda\kappa' - \Phi\kappa' \kappa_0^T\Phi_r^T \Phi_r\Lambda_r\kappa_0)\\ \nonumber
 &=\kappa_0^T \Phi_r^T \Phi (\Pi \Lambda\kappa' -
\kappa'\kappa_0^{-1} \Lambda_r\kappa_0)\kappa_0^{-1} \kappa_0\\\nonumber
&=\kappa_0^T \Phi_r^T \Phi(\Pi \Lambda\rho' -
\rho' \Lambda_r) \kappa_0\\\nonumber
&=\kappa_0^T \Phi_r^T\begin{pmatrix}
 \Phi_r  & \Phi_{r^c} \\
\end{pmatrix} (\Pi \Lambda\rho' -
\rho' \Lambda_r) \kappa_0\\\nonumber
&=\kappa_0^{-1} \begin{pmatrix}
 I_{r\times r}  &  \Theta_{r\times r_c} \\
\end{pmatrix} (\Pi \Lambda\rho' -
\rho' \Lambda_r) \kappa_0.
   \end{align*}
 Therefore, the  orthogonality constraint
    $\partial U'(t)/\partial t \perp U_0 $ is equivalent to:
  \begin{equation}\label{orth_const_aux}
  \begin{pmatrix}
 I_{r\times r}  &  \Theta_{r\times r_c} \\
\end{pmatrix} (\Pi \Lambda\rho' -
\rho' \Lambda_r) = 0.
  \end{equation}
   It follows that
    \begin{equation}
  \begin{pmatrix}
 I_{r\times r}  &  \Theta_{r\times r_c} \\
\end{pmatrix} \begin{pmatrix}
 0 & \Theta\Lambda_{r^c}   \\
 0 & \Lambda_{r^c}   \\
\end{pmatrix} \begin{pmatrix}
 \rho'_r \\
 \rho'_{r_c}
\end{pmatrix} - \begin{pmatrix}
 I_{r\times r}  &  \Theta_{r\times r_c} \\
\end{pmatrix} \begin{pmatrix}
 \rho'_r \Lambda_{r} \\
 \rho'_{r_c} \Lambda_{r}
\end{pmatrix} = 0.
  \end{equation}
  or equivalently,
  \begin{equation}\label{ortho_cont_aux_2}
  \Theta\Lambda_{r_c}\rho'_{r_c}-\rho'_r\Lambda_r=\Theta(\rho'_{r_c}\Lambda_r -\Lambda_{r_c}\rho'_{r_c}  ).
  \end{equation}
  Using the orthogonality constraint given by equation (\ref{ortho_cont_aux_2}), and using equation (\ref{evol_perturb_aux}), the evolution equation for $\rho'_r(t)$ becomes:
\begin{equation}
\frac{d \rho_r'}{dt}=\Theta(\rho'_{r_c}\Lambda_r
-\Lambda_{r_c}\rho'_{r_c}  ).
\end{equation}
The above equation shows that the evolution of $\rho_r'(t)$ can be expressed solely based  on the evolution  of $\rho_{r_c}'(t)$. Thus, the stability of the solution only depends on the stability of  $ \rho'_{r_c}$.
The evolution of $ \rho'_{r_c}$ in the index notation is given by:\begin{equation}
\frac{d \rho'_{ij}}{d t} = (\lambda_i -\lambda_j) \rho'_{ij}, \quad \quad i=r+1, r+2, \dots, n,  \quad \quad 
\mbox{and} \quad \quad j=1, 2, \dots, r.
    \end{equation}
    Therefore, the asymptotic stability of $ \rho'$ requires that:
\begin{equation}
\mbox{real}(\lambda_i) \leq \mbox{real}(\lambda_j), \quad \quad \mbox{for} \quad \quad i=r+1, r+2, \dots, n,  \quad \quad 
\mbox{and} \quad \quad j=1, 2, \dots, r.
\label{stability_cond}
\end{equation}
 The inequality  (\ref{stability_cond}) shows that the subspace $\Phi_r$ with eigenvalues having the largest real part is the only stable solution to equation (\ref{dynamical_sys_sub}). This completes the proof.
\subsection{Time-dependent linearized dynamics\\ }
Consider the evolution of an arbitrary (not orthonormal)  subspace $V(t) \in \mathbb{R}^{n\times r}$ under the time-dependent linearized dynamics which is governed by:
\begin{align}\label{eq:time_dep_lin}
\frac{\partial V}{\partial t}&=L(t)V,\\ \nonumber
V(0)&=V_0,
\end{align}
and consider the corresponding OTD evolution:
\begin{align}\label{eq:time_dep_otd}
\frac{\partial U}{\partial t}&=L(t)U-UL_r(t),\\ \nonumber
U(0)&=U_0,
\end{align}
We choose the initial state of the OTD basis such that $U_0$ and $V_0$ span the  same subspace. In the following Theorem  we  show that the OTD modes exactly span the  subspace  $V(t)$. More specifically, we show that the two subspaces $U(t)$ and $V(t)$ are related via   a time-dependent transformation matrix. 
\begin{theorem}\label{Thm:Trans}
 Let   $V(t) \in \mathbb{R}^{n\times r}$ be the subspace evolved under the time-dependent linearized dynamics, eq. (\ref{eq:time_dep_lin}); and $U(t) \in \mathbb{R}^{n\times r}$ be the OTD basis evolved with eq. (\ref{eq:time_dep_otd}).
Then, assuming that initially the two subspaces are equivalent, i.e. there is a matrix $T_0$ such that $V_0 = U_0T_0$,  there exists a linear transformation $T(t)$ such that $V(t) = U(t)T(t),$ where $T(t)$ is the solution of the matrix differential equation:
\begin{align}\label{eq:Tdot}
\frac{dT(t)}{dt}&=L_r(t)T(t)\\ \nonumber
T(0)&=T_0.
\end{align}
\end{theorem}
\textbf{Proof:}
We substitute the transformation $V(t)=U(t)T(t)$ into the evolution equation  for $V(t)$:
 \begin{align*}
 \dot{U}T+U\dot{T}=LUT
 \end{align*}
 Substituting $\dot{T}$ from equation (\ref{eq:Tdot}) after rearrangement results in:
\begin{align*}
 \dot{U}T=LUT-UL_rT.
 \end{align*}
 Multiplying both sides of the above equation by $T^{-1}$ from the right results in:
\begin{align*}
 \dot{U}=LU-UL_r,
 \end{align*}
 which is the OTD equation for $U(t)$. The initial condition is also $U(0)=U_0$. This completes the proof.

For dynamical systems with persistent instabilities the evolution  under the time-dependent linearized equation (\ref{eq:time_dep_lin}) is unstable, and $V(t)$ grows rapidly. Even for stable dynamical systems, as we move away from $t=0$, almost any vector approaches to the least stable direction. However the evolution of the OTD modes, due to their built-in orthonormality, is always stable, and as we will  demonstrate in our results, the OTD evolution leads to a well-conditioned numerical algorithm that peels off the most unstable directions of the dynamics. 
\subsection{Mode ranking within the subspace}
Having derived the subspace that corresponds to the most unstable directions, the next step is to rank these directions internally, i.e. within the subspace. As we describe below, there are two ways to rank the OTD basis  based  on the  growth rate. Both of these approaches amount to   an internal rotation within the OTD subspace.

(1) The  \emph{instantaneous growth rate} $\sigma_i(t)$ \cite{farrell_2, Haller2010} is obtained by computing
the eigenvalues of the symmetric part of $L_r$, \emph{i.e} $L_r^{\sigma} =(L_r+L_r^T)/2$:
\begin{equation}\label{eq:Inst_grth}
L_r^{\sigma} R^{\sigma} = R^{\sigma} \Sigma,
\end{equation} 
where $\Sigma=\mbox{diag} (\sigma_1(t), \sigma_2(t), \dots, \sigma_r(t)
)$, and $R^{\sigma}(t) \in \mathbb{R}^{r\times r} $ is the  matrix of eigenvectors.
We rank these values from the least stable to the most stable directions,
\emph{i.e.}:
\begin{equation*}
\sigma_1 (t)  \geq \sigma_2 (t)  \dots  \geq \sigma_r(t).   
\end{equation*}
We note that $\sigma_{\max}(t) =\sigma_1(t)$ is the \emph{numerical abscissa} of the operator $L_r$.
Therefore $\sigma_{\max}(t)$ represents the maximum instantaneous growth rate within the OTD subspace.

(2) The \emph{instantaneous eigenvectors} of the reduced operator can also be obtained through the equation:
\begin{equation}\label{eq:s2}
L_rR^{\lambda} = R^{\lambda} \Lambda_{r},
\end{equation} 
where the eigendirections are represented by   $R^{\lambda}(t) \in  \mathbb{C}^{r\times r} $. The  instantaneous eigenvalues are denoted by
$\Lambda_r(t)  = \mbox{diag} (\lambda_1(t), \lambda_2(t), \dots, \lambda_r(t) )$, where $\lambda_i$ are ranked from the eigenvalue with the largest real part $\lambda_1(t)$, to the eigenvalue with the smallest real part $\lambda_r(t)$. 

Based on the above two rankings we define the rotated OTD basis:
 \begin{equation}
U^{\lambda,\sigma}(t) = U(t)R^{\lambda,\sigma}(t),  
 \end{equation} 
where $U^{\lambda,\sigma}(t) \in \mathbb{C}^{n\times r}$ is the ranked representation of the OTD modes defining the space $U$, and $R^{\lambda,\sigma}$ is the rotation matrix obtained from either of the two described strategies for mode ranking (eq. (\ref{eq:Inst_grth}) or eq. (\ref{eq:s2})).
   For a non-Hermitian
operator ${L}$,  the ranking based on the instantaneous eigenvectors $R^{\lambda}(t),$ results in  modes $U^{\lambda,\sigma}$ which many not be mutually orthogonal.
For this case, the orthogonal 
representation of the least stable directions can be obtained by performing
a Gram-Schmidt orthonormalization.

\section{Linear dynamics}
Linear dynamics and in particular non-normal behaviour has a critical role in determining the short-time fate  of a disturbance in both linear and nonlinear dynamical systems. To study the evolution of  an initial condition under the effect  of
linear or linearized dynamics, reduction in eigenfunction coordinates $U^{\lambda,\sigma}$  is often utilized. However for highly non-normal operators, a large number of eigenfunctions are required to correctly capture the non-normal behavior, since the eigenfunctions are far from  orthogonal  and, in fact they constitute a highly ill-conditioned basis (\emph{i.e.} $\|U^{\lambda,\sigma}\|\|(U^{\lambda,\sigma})^{-1}\|\gg1).$  In what follows we demonstrate the computational efficiency of the OTD modes in  capturing non-normal behavior  and   contrast the OTD basis with  the eigenfunction basis for linear systems.  
\subsection{Advection-diffusion operator}
As the first case, we consider the advection-diffusion operator which has
a wide range of applications in fluid mechanics, financial mathematics,
and many other fields. Particularly we are interested to study the effect
of non-normality on the reduced operator, both in the short-time and the long-term
asymptotic behavior.
The operator with zero boundary condition is given by:
\begin{align}\label{Adv-Diff}
{L}(u) &= \nu \frac{\partial^2 u}{\partial x^2} + c \frac{\partial
u}{\partial x}, \quad \quad x \in [0, 1] \\ \nonumber
u(0,t) &= 0, \quad  u(1,t) = 0. 
\end{align}
where $\nu$ is the diffusion coefficient  and $c$ is the advection speed.
The evolution equation for the OTD basis is expressed by:
\begin{align}\label{Adv-Diff}
\frac{\partial u_i}{\partial t}&={L}(u_i)-\big<{L}(u_i),u_j\big>u_j, \quad \quad i,j,=1,\dots, r. \\ \nonumber
u_i(0,t) &= 0, \quad  u_i(1,t) = 0, 
\end{align}
where the inner product and the induced norm are:
\begin{equation*}
\big< u,v\big> := \int_0^1 u(x)v(x)dx, \quad \quad \|u\|:= \big< u,u\big>^{1/2}.
 \end{equation*}

To solve the evolution equation  we use Chebyshev collocation discretization
implemented in 
 \emph{chebfun} \cite{chebfun}. The OTD basis is initialized with orthonormalized
modes:
 \begin{equation*}
u_n(x,0) = \frac{\sin(n \pi x)}{\| \sin(n \pi x)\|}, \quad n=1,2,\dots, r,
 \end{equation*}
 for all the cases considered in this section.

In the case of a nearly normal operator, \emph{i.e} large $\nu$, an optimal
basis must be close to the dominant eigenfunctions for all times. On the
other hand,
for a set of parameters that corresponds to a non-normal operator, an optimal basis should differ from the eigenfunctions
for short-time dynamics, and only for $t \gg\ 1$  should it converge to the operator
eigenfunctions. In the remaining of this section, we demonstrate that the
OTD subspace captures short- and long-time dynamics for both normal and
non-normal operators.

\begin{figure}
\centering
\subfigure[]{
\includegraphics[]{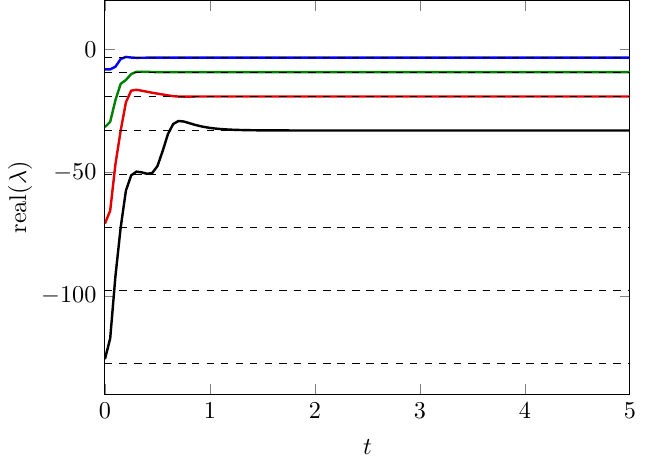}
}
\subfigure[]{
\includegraphics[]{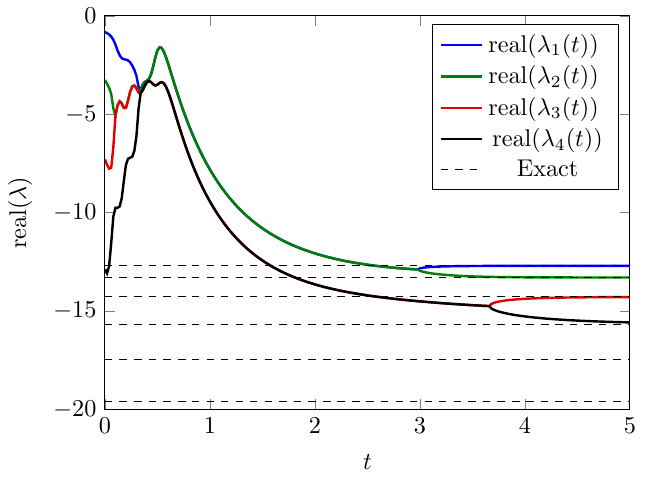}
}
\caption{Instantaneous eigenvalues of the reduced operator with $r=4$ for
$c=1$. The dashed lines show the  eigenvalues of $\mathcal{L}$ with the
largest real part. (a) $\nu=0.2$; (b) $\nu=0.02$.   }
\label{fig_eigvl_rope}
\end{figure}
\begin{figure}
\centering
\subfigure[]{
\includegraphics[]{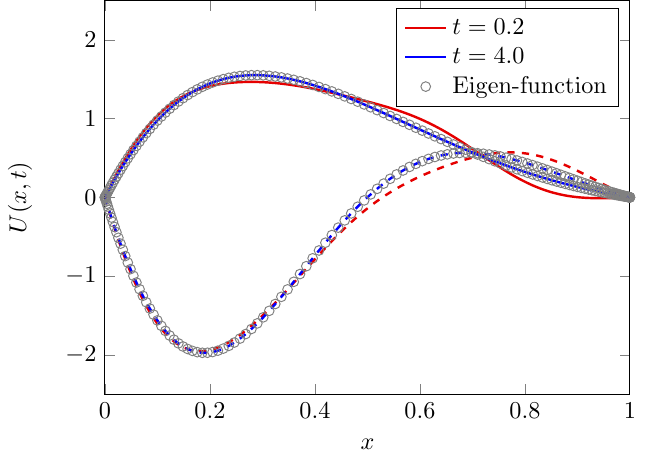}
}
\subfigure[]{
\includegraphics[]{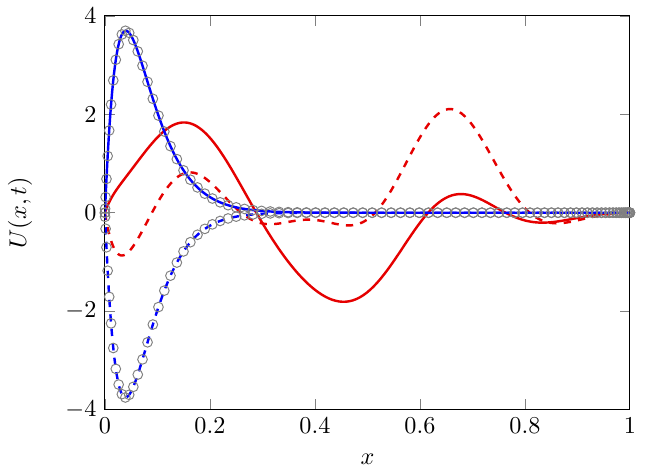}
}
\caption{OTD basis for $r=4$ and $c=1$. The circles show the first two dominant eigenfunctions
of ${L}$. The solid and dashed lines show the first and second OTD modes, \emph{i.e.}
$u^{\lambda}_1(x,t)$ and $u^{\lambda}_2(x,t)$:
 (a) $\nu=0.2$; (b) $\nu=0.02$.   }
 \label{fig_modes_adv}
\end{figure}

To  analyze the behavior of the OTD basis we choose  $r=4$.  The instantaneous
eigenvalues  of the reduced operator, $\mbox{real}(\lambda_i(t))$, for $\nu=0.2$ are shown with colored solid lines 
in  Figure \ref{fig_eigvl_rope}(a).
The  real part of the   least stable eigenvalues of  the operator ${L}$ are shown with the  dashed lines. It is clear that the instantaneous eigenvalues  quickly approach
the least stable eigenvalues of ${L}$.  In Figure \ref{fig_modes_adv}(a),
the first two elements of the OTD basis, internally rotated in the eigendirection, $\phi_1$ and $\phi_2$, are shown at
$t=0.2$ and $t=4.0$. These are superimposed with the two most unstable eigenfunctions of the operator. At $t=0.2$, the modes are very close to the
eigenfunctions
of ${L}$  and at $t=4.0$ the modes have essentially converged to
the eigenfunctions. This demonstrates that due to the strongly normal behavior
of the operator at 
$\nu=0.2$, the eigenfunctions explain the dynamics accurately for both short
time and long time, and the OTD basis quickly converges to the space
spanned by the eigenfunctions. 

At $\nu=0.02$, however, the instantaneous eigenvalues converge with a much
slower rate and much later, \emph{i.e.} $t>3$,  to the operator eigenvalues,
as it is shown
in Figure \ref{fig_eigvl_rope}(b). Accordingly, as it can be seen in Figure
\ref{fig_modes_adv}(b), the OTD basis elements are different from the eigenfunctions
of
 ${L}$ at $t=0.2$, and only until later the basis approaches the eigenfunctions.
Another important observation is related with the advection direction of the OTD basis, which is left-ward. For instance at $t=0.2$, the basis has advected an approximate distance of $\Delta x = c \Delta t= 0.2$ to the left. As a result the OTD basis
 has a strong presence in the region $0< x <0.8$. This is not the case for the eigenfunctions
of ${L}$ which are primarily concentrated near $x=0$.   

\subsection{Orr-Sommerfeld/Squire operator}
As the second case we consider the Orr-Sommerfeld/Squire (OS/SQ) equation that governs the evolution of infinitesimal disturbances in parallel viscous flows.  The eigenvalues of OS/SQ operator are  highly sensitive to perturbations, and its eigenfunctions are linearly dependent, resulting in a highly  ill-conditioned linear dynamical system.  To this end, the OS/SQ equation is considered only for demonstration purposes, i.e. to illustrate that the OTD modes capture correctly the short-time evolution of the infinitesimal disturbances, as well as their asymptotic (long-term) behavior.  

We consider the plane Poiseuille flow in which the base-flow velocity is uni-directional given by $\mathbf{u}(x,y,z) = U(y) \mathbf{e}_x$, with $U(y) = 1-y^2$. 
The disturbance is taken to be:
\begin{equation*}
\mathbf{v}'(x,y,z,t) = \mathbf{v}(y,t) \exp(i \alpha x + i\beta z),
\end{equation*}
with 
\begin{equation*}
\mathbf{v'} =
 \begin{pmatrix}
  v'    \\
 \eta'  \\
 \end{pmatrix},
 \quad \quad \mbox{and} \quad
 \mathbf{v} =
 \begin{pmatrix}
  v    \\
 \eta  \\
 \end{pmatrix},
\end{equation*}
where $v'$ and $\eta'$ are the wall-normal velocity and the vorticity, respectively, and $\alpha$ and $\beta$ denote the streamwise and spanwise wave numbers, respectively.

The 
Orr-Sommerfeld equation in velocity-vorticity formulation is given by:
\begin{equation}\label{OS-Compact}
\frac{\partial \mathbf{v}}{\partial t} = L \mathbf{v},
\end{equation}
with boundary conditions:
\begin{equation}
v = \mathcal{D}v=\eta=0 \quad \quad \mbox{at} \quad \quad y=\pm 1, 
\end{equation}
where $L$ is a linear operator:
\begin{equation}
L =
 \begin{pmatrix}
  L_{OS} & 0   \\
  L_{C}  & L_{SQ}  \\
 \end{pmatrix},
\end{equation}
with:
\begin{align*}
L_{OS} &= (\mathcal{D}^2 - k^2)^{-1} \bigg[\frac{1}{Re} (\mathcal{D}^2 - k^2)^2  + i\alpha \mathcal{D}^2 U - i\alpha  U (\mathcal{D}^2 - k^2) \bigg],\\
L_{C} &= -i\beta \mathcal{D} U,\\
L_{SQ} &= \frac{1}{Re} (\mathcal{D}^2 - k^2)- i\alpha \mathcal{D} U,
\end{align*}
and $k =\sqrt{\alpha^2+\beta^2}$ is the modulus of the wave vector and $\mathcal{D} = \partial/\partial y$.
For the complete derivation of the OS/SQ equations,  we refer to \cite{Henningson_Schmidt}.

For the choice of the inner product we use the energy measure, which provides a physically motivated norm that arises naturally from the OS/SQ equation \cite{schmid2007nonmodal}. The inner product is given by:
\begin{equation}
\big < \mathbf{v}_1, \mathbf{v}_2\big >_E:=\frac{1}{k^2} \int_{-1}^1 \mathbf{v}_1^H M \mathbf{v}_2  dy,
\end{equation} 
where:
\begin{equation}
M =
 \begin{pmatrix}
  k^2 - \mathcal{D}^2  & 0   \\
  0  & 1 \\
 \end{pmatrix},
\end{equation}
and $( \ )^H$ denotes complex conjugate. In the following we consider a discrete representation of the operator $L \in \mathbb{C}^{n \times n}$.  Assuming a solution of the form:
\begin{equation*}
\mathbf{v}=\boldsymbol \phi \exp{\lambda t,}
\end{equation*}
the initial value problem (\ref{OS-Compact}) transforms to an eigenvalue
problem of the form
\begin{equation*}
L \Phi=  \Phi \Lambda,
\end{equation*}
where $\Lambda = \mbox{diag}(\lambda_1, \lambda_2, \dots, \lambda_n) $ and $\Phi=\{\phi_1 | \phi_2 | \dots |\phi_n\}$ are, respectively, the matrices of the eigenvalues and the eigenvectors  of $L$. 

Orszag \cite{orszag1971accurate} showed that for $Re \leq Re_c \simeq 5772.22$ all  eigenvalues of the operator $L$ have negative real parts and
 therefore any perturbation is asymptotically stable. However, even for $Re < Re_c$, the energy of a perturbation may experience significant transient growth. 
 This is a direct consequence of the non-normality of $L$. In this section we look at the evolution of the  OTD modes for the OS/SQ operator.
 In particular we consider a three-dimensional perturbation with $\alpha=1$ and $\beta=1$ at $Re = 5000$ that  corresponds to an asymptotically stable operator.

Since the dynamical system is linear, the linear tangent operator and  $L$
are identical.  Thus, the evolution equation for the OTD modes $\mathbf{U}=\{\mathbf{u}_1,\mathbf{u}_2, \dots, \mathbf{u}_r \}$ becomes:
\begin{equation}
\frac{\partial \mathbf{u}_i}{\partial t} = L \mathbf{u}_i - \big
< L \mathbf{u}_i , \mathbf{u}_j\big >_E \mathbf{u}_j,  \quad \quad \mathbf{u}_i=
 \begin{pmatrix}
  v_i    \\
 \eta_i  \\
 \end{pmatrix}, \quad \quad
i,j=1,2, \dots, r,
\end{equation}
with the boundary conditions:
\begin{equation}\label{BC_OS_OT}
 v_i = \mathcal{D}v_i=\eta_i=0 \quad \quad \mbox{at} \quad \quad y=\pm 1 \quad
\quad i=1,2, \dots, r.
\end{equation}
For  space   we use Chebyshev collocation discretization
with 256 points, while for time advancement we  use the  first order implicit Euler method.
\subsubsection{Initial condition}\label{sec:OS_IC}
We initialize the OTD subspace such that its span encompasses the \emph{optimal initial condition}: an initial condition $\mathbf{v}^0$ that reaches the maximum 
possible amplification at a given time $t=t_{\mbox{max}}$. The optimal initial condition can be formulated as \cite{schmid2007nonmodal}:
 \begin{equation}
G(t_{\mbox{max}})  = \underset{\mathbf{v}_0 \neq 0}{\mbox{max}} \frac{\| \mathbf{v}(t_{\mbox{max}}) \|^2_E}{\| \mathbf{v}_0 \|^2_E} = \| \exp(  \Lambda t_{\mbox{max}}) \|^2_E
\end{equation} 
The value of $\| \exp(  \Lambda t_{\mbox{max}}) \|^2_E$ is equal to the  principal singular value $s_1$ of the propagator $\mathbf{\Phi}_{t_{0}}^{t_{\mbox{max}}}=\exp(  L t_{\mbox{max}})$.
It follows that:\begin{equation}\label{optimal_init}
\mathbf{\Phi}_{t_{0}}^{t_{\mbox{max}}} \mathbf{V}_0 = \mathbf{V}\mathbf{S}, 
\end{equation} 
where $\mathbf{V}(t_{\mbox{max}})=\{\mathbf{v}_1(t_{\mbox{max}}),...,\mathbf{v}_n(t_{\mbox{max}})\}$ are the right singular  eigenvectors  and $\mathbf{V}_0=\{\mathbf{v}_{1_0},\mathbf{v}_{2_0},...,\mathbf{v}_{n_0}\}$ 
are the left singular  eigenvectors, and $\mathbf{S} = \mbox{diag} \{s_1, s_2, \dots  ,s_{n}\}$ consists of  singular  values of the operator $B$. The initial 
state of the subspace of size $r$ is chosen to be:
\begin{equation*}
 \mathbf{u}_{i_0}=\mathbf{v}_{i_0}, \quad i=1,\dots, r.
 \end{equation*}
  The admissible initial conditions for the OTD modes must satisfy (i) the orthonormality constraint, and (ii) the boundary conditions at the walls given by equation (\ref{BC_OS_OT}). It is straightforward to show that the above choice is compatible with these criteria. We also note that short of  these criteria, the choice of initial conditions for the OTD\ subspace is arbitrary.  Certainly, the choice of optimal initial condition is of high practical importance with significant attention  paid to in the literature. We refer the readers to \cite{schmid2007nonmodal} and references therein. Moreover,  due to the non-normality of the operator, the optimal initial condition  requires a large number of eigenmodes for accurate representation, resulting in a relatively high-dimensional system in eigenmode coordinates compared with the OTD modes.  
\subsubsection{Transformation matrix}
We  obtain a time-dependent reduction of the OS/SQ operator by projecting $L$ onto  the OTD subspace using the energy inner product:
\begin{equation}
L_{r_{ij}}(t)= \big <\mathbf{u}_i,  L(\mathbf{u}_j) \big >_E \quad i,j=1,
\dots, r,
\end{equation}
where $L_r(t)\in \mathbb{C}^{r\times r}$ is the reduced OS/SQ operator. The reduced operator is then used to evolve the transformation matrix $T(t) \in \mathbb{C}^{r \times r}$ according to equation (\ref{eq:Tdot}).  
Using the same initial condition $\mathbf{V}_0$ for both OTD modes and OS/SQ, results in $T_0$ being the identity matrix, \emph{i.e} $T_0=I$. In the following section we compare the evolution of $\mathbf{V}_0$ under the full OS/SQ operator with the evolution of $\mathbf{V}_0$ using the transformation relation $\mathbf{V}(t)=\mathbf{U}(t)T(t)$.

\subsubsection{Asymptotically stable subspace}
\begin{figure}
\centering
 \includegraphics[width=0.7\textwidth]{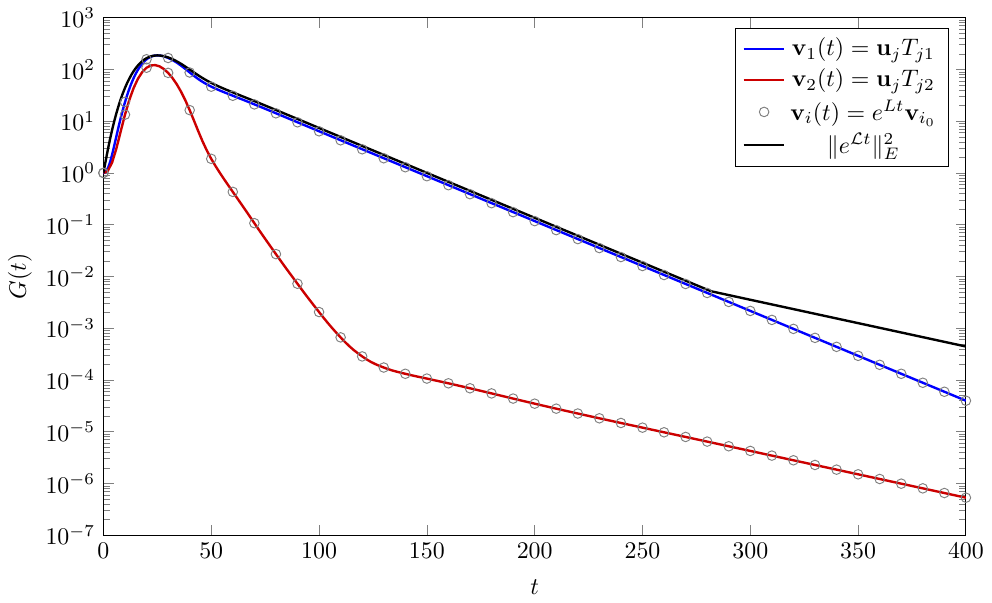}
\caption{Transient energy growth, $G(t)= \| \mathbf{v} \|^2_E$ for plane
Poiseuille flow  at $Re=5000$ with $\alpha=1$ and $\beta=1$.  The solid lines
 (blue and red) show the energy growth of two different initial perturbations
computed with the reduced operator with $r=2$. The  circles show the exact energy growth
computed with the OS/SQ operator.}
\label{OS_Tran_Grth_Re=5000}
\end{figure}

We   consider the evolution of the  OTD subspace with $r=2$  and $r=3$ for three-dimensional perturbations  at $Re=5000$ and streamwise and spanwise wave numbers of   $\alpha=1$ and $\beta=1$. At this Reynolds number all perturbations are asymptotically stable, while some perturbations  experience significant non-normal growth in the short-time regime. In Figure \ref{OS_Tran_Grth_Re=5000}, the norm of the solution operator, $\|e^{Lt}\|_E^2$,  is shown. As it can be seen, the energy of some initial conditions is amplified by  a factor of  over one hundred. The maximum energy growth can be achieved at $t_{\mbox{max}}=25.06$ for the optimal initial condition.  The optimal initial condition is obtained from equation (\ref{optimal_init}). We  initialize the two OTD modes with the first two elements of the right singular eigenvectors $\mathbf{V}_0$. 

\begin{figure}
\centering
 \includegraphics[width=0.7\textwidth]{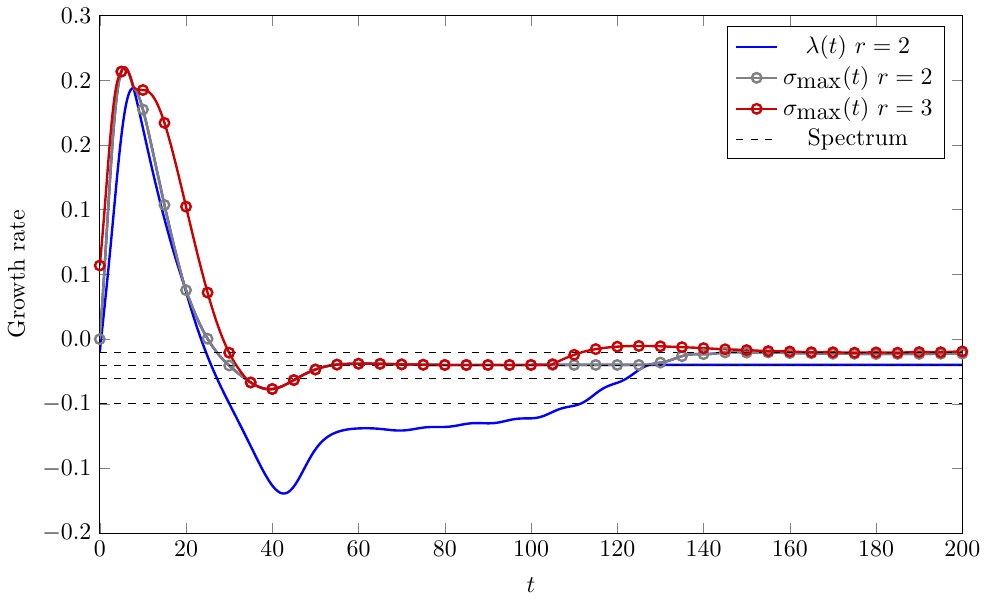}
\caption{ Plane Poiseuille
flow  at $Re=5000$, $\alpha=1$ and $\beta=1$: instantaneous real eigenvalues with $r=2$ (blue lines); numerical abscissa  with $r=2$ (gray lines);  numerical abscissa  with $r=3$ (red);  the real part of the first four least stable  eigenvalues
of the OS/SQ operator (dashed lines).}
\label{OS_Inst_Grth_Rate_Re=5000}
\end{figure}

Now, we first compare  the evolution of the initial subspace with $r=2$ using the reduced
operator with that of the  full OS/SQ
operator for the choice of initial condition explained in Section \ref{sec:OS_IC}.
In Figure \ref{OS_Tran_Grth_Re=5000}, we compare the energy of $\mathbf{v}_1(t)$
and $\mathbf{v}_2(t)$ obtained from   (i) evolution of the OTD and using the transformation matrix $T(t)$ to obtain $\mathbf{v}_i(t)=\mathbf{u}_j(t)T_{ji}(t)
,\ i,j=1, \dots, r$, and (ii) by solving the full OS/SQ operator, \emph{i.e.} $\mathbf{v}_i(t) = \mathbf{\Phi}_{0}^{t}\mathbf{v}_{i_0}=e^{L t}\mathbf{v}_{i_0}$. 
In both cases, excellent agreement in both short-time and large-time evolution
is observed. This demonstrates that the OTD modes  correctly follow the evolution
of a class of initial conditions according to Theorem \ref{Thm:Trans}. Given that at $Re=5000$ the OS/SQ operator
is highly non-normal, a large number of eigenmodes are required to correctly
follow the evolution of an initial condition. However the OTD modes, do not
require additional modes beyond the  dimension of the initial subspace.
In Figure  \ref{OS_Inst_Grth_Rate_Re=5000}  the instantaneous eigenvalues  for $r=2$ along
with numerical abscissa for $r=2$ and $r=3$ are shown. The black dashed lines  show the real  parts of the eigenvalues of the OS/SQ operator. The eigenvalues shown  are the four most unstable ones  of the OS/SQ operator. The significant non-normal energy growth manifests itself with  positive real eigenvalues and instantaneous growth rates in finite time, despite all eigenvalues of OS/SQ having negative
real parts. The instantaneous eigenvalues for the case with $r=2$   converge to the first two least stable eigenvalues of the OS/SQ operator in accordance to Theorem \ref{Thm:SLD}. For $r=2$, the largest real instantaneous eigenvalue and the numerical abscissa $\sigma_{\mbox{max}}$ are nearly identical. This shows that $\mathbf{u}_1(t)$ is nearly aligned with the direction of maximum growth for all times.  
\begin{figure}[!t]
\centering
\includegraphics[width=.85\textwidth]{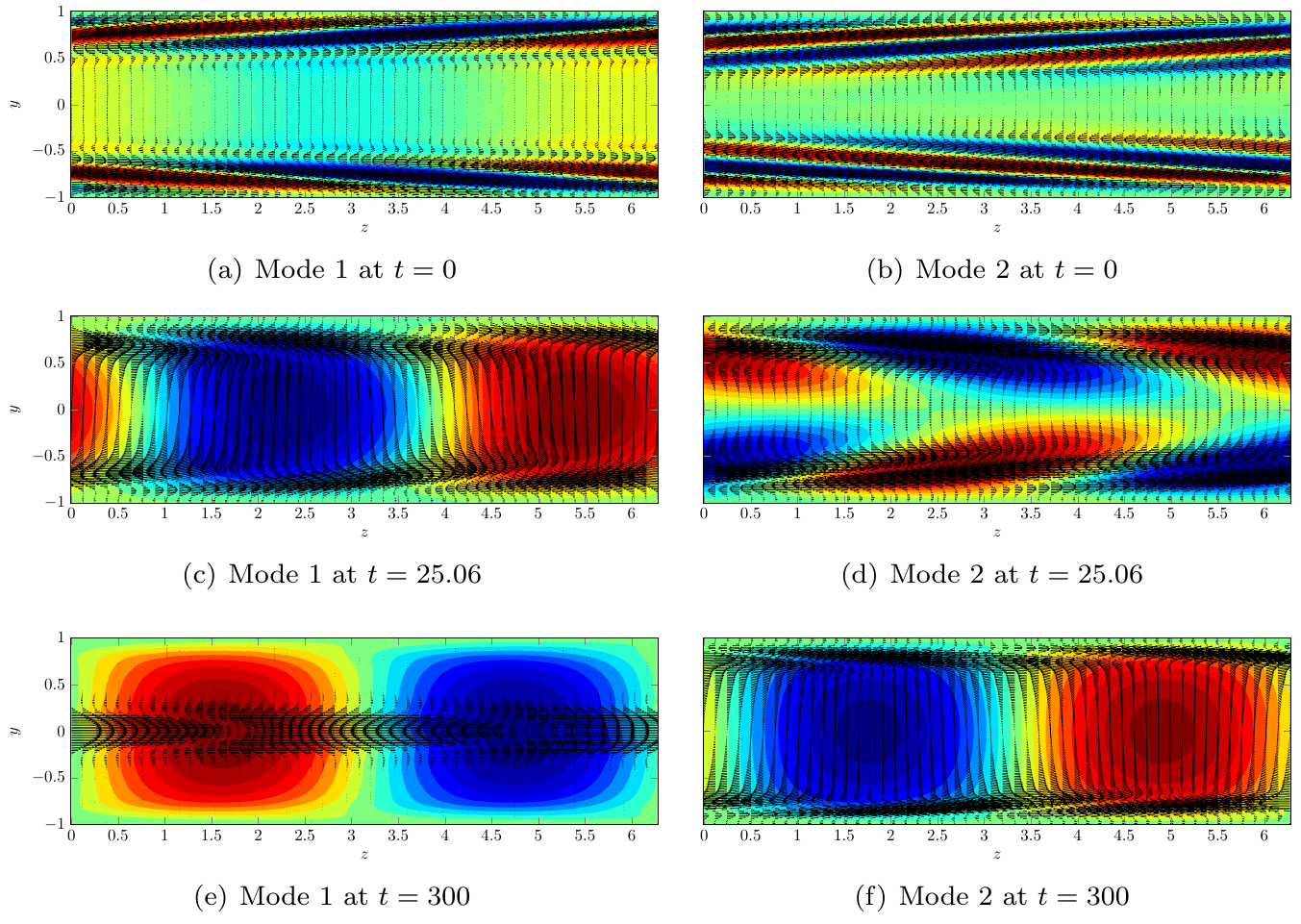}
\caption{Snapshots of the OTD modes with $r=2$  at $Re=5000$ and $\alpha=1$ and $\beta=1$ in the streamwise plane $x=\pi$. The contour shows the vertical velocity.  } 
\label{Modes-Re=5000-a=1-b=1}
\end{figure}
Now we explore some  aspects of increasing the dimension of OTD from $r=2$ to  $r=3$. For the sake of brevity, let us denote  the  quantities for the case $r=3$ with the superscript $( \ )'$. The initial condition of the cases with $r=2$ and $r=3$ are $\mathbf{V}_0=\{\mathbf{v}_{1_0},\mathbf{v}_{2_0}\}$
and $\mathbf{V}'_0=\{\mathbf{v}_{1_0},\mathbf{v}_{2_0},\mathbf{v}_{3_0}\}$
respectively. Clearly $\mathbf{V}_0  \subset  \mathbf{V}'_0$ and consequently  $\mathbf{V}(t) \subset \mathbf{V}'(t)$ for all times.
 From the transformation between $\mathbf{U}(t)$ and $\mathbf{V}(t),$ we can deduce that the embedding of the initial condition is preserved for the OTD subspaces as well, \emph{i.e.} $\mathbf{U}(t) \subset \mathbf{U}'(t)$ for all times. 
Therefore it is to be expected that the maximum growth rate in the case of $r=3$ must be always larger  (or equal) than the corresponding value in the case of  $r=2$. In other words  
 $\sigma_{\mbox{max}}(t) \leq \sigma'_{\mbox{max}}(t) $ for all times. This observation is confirmed in comparing the numerical abscissa for $r=2$ and $r=3$ as shown in Figure \ref{OS_Inst_Grth_Rate_Re=5000}.  The abrupt changes in the values of numerical abscissa are the result of eigenvalue crossing in the symmetric part of the reduced operator $L_r$, where the direction of maximum growth switches from one eigendirection to the other. 

Figure \ref{Modes-Re=5000-a=1-b=1} shows the OTD modes for the case of $r=2$ at three time instants of their evolution at the streamwise plane $x=\pi$: (i) the initial state $t=0$, (ii) maximum energy  $t=t_{\mbox{max}}$, and (iii) at large time $t=300$. The initial state is marked by flow patterns that oppose the base shear.
As times evolves from $t=0$ to $t<t_{\mbox{max}}$, the OTD modes tilt into the mean shear flow, resulting in significant growth rates for the subspace. At $t=300$ the modes eventually approach the two most unstable eigenmodes of the OS/SQ operator. This demonstrates that the time-dependent  modes capture characteristically different regimes in the evolution of the subspace.

\section{Nonlinear dynamics }
Here we consider two nonlinear systems for which we compute the OTD modes. The first system is a low-dimensional dynamical system, while the second one is a more complex application involving an unstable flow with strongly transient characteristics. 
\subsection{Low-dimensional dynamical system}
\begin{figure}
\centering
\subfigure[]{
\includegraphics[width=.48\textwidth]{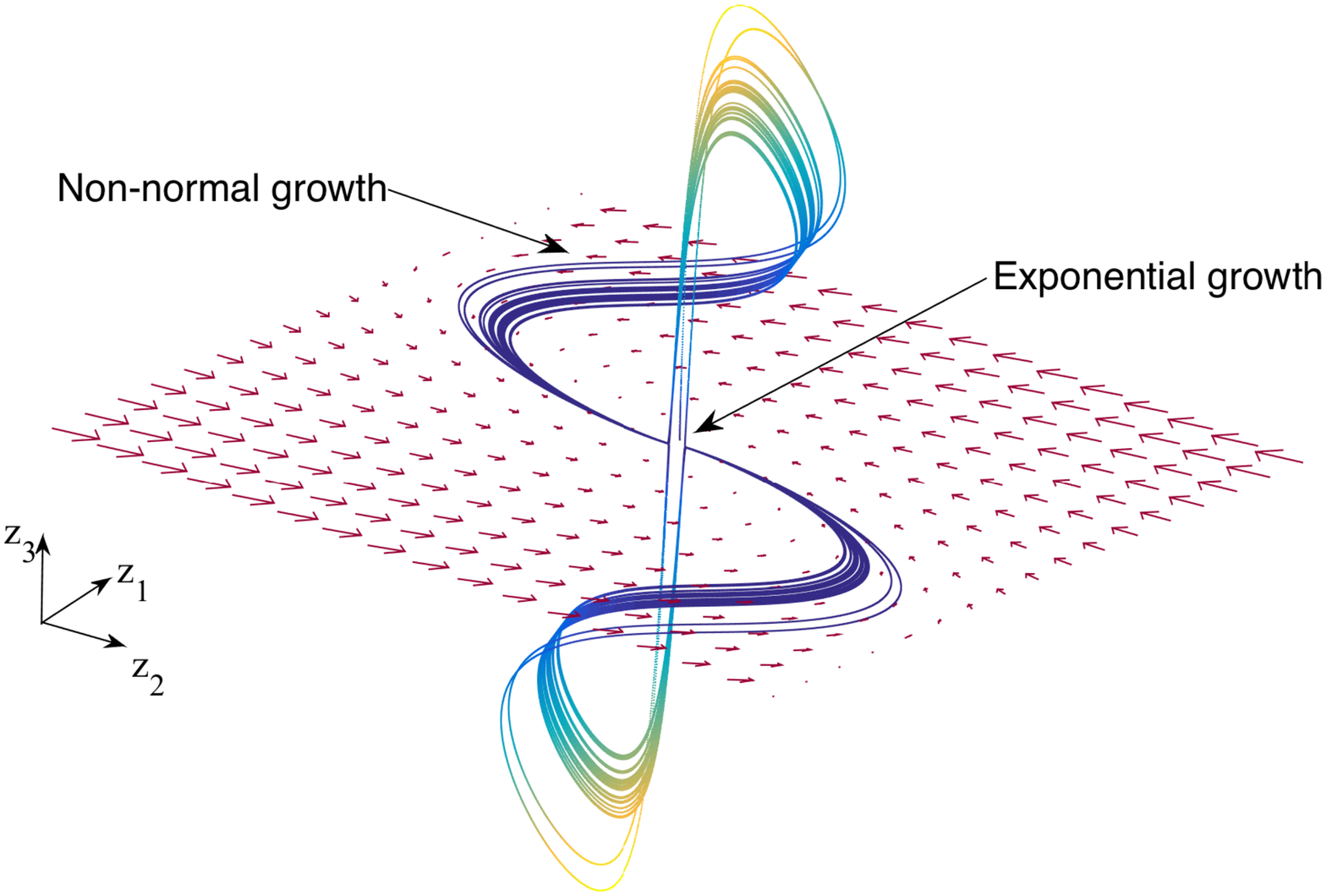}
}
\subfigure[]{
\includegraphics[width=.43\textwidth]{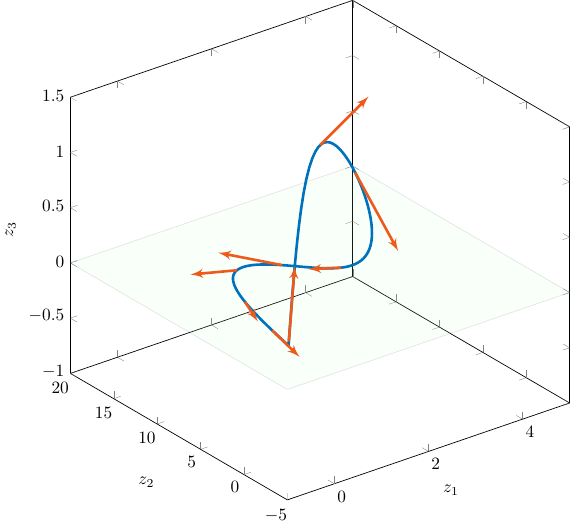}
}
\subfigure[]{
\includegraphics[width=.43\textwidth]{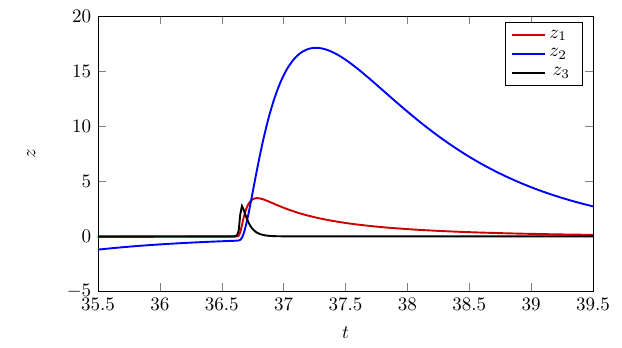}
}
\subfigure[]{
\includegraphics[width=.43\textwidth]{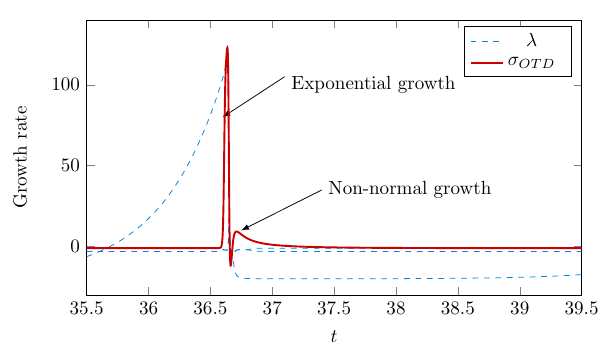}
}
\caption{(a) A trajectory of the considered dynamical system colored according to the state variable $z_3$. The non-normal vector field for $z_3=0$ is also shown; (b) A single cycle of the trajectory is shown together with a single OTD mode; (c) The time series for $z_i(t), i=1,2,3$; (d) The three eigenvalues of the linearized operator are plotted with blue dashed curves while the growth rate of the single OTD mode is shown with red color.} 
\label{fig_low}
\end{figure}

We design a low-dimensional dynamical system in order to demonstrate  transient growth over different directions and how these can be captured by the developed approach. In particular we consider the following  system:
\begin{align}
\frac{dz_1}{dt}&=-a_1z_1 +\epsilon z_2+bz_3\\
\frac{dz_2}{dt}&= \epsilon^{-1} z_1 -a_2z_2\\
\frac{dz_3}{dt}&=bz_3(\frac{1}{\sqrt{z_1^2+z_2^2}}-1),
\end{align}
where we take $a_1=a_2=2$, $\epsilon=0.05$, and $b=20$.
For these parameters the dynamical system has an almost periodic behavior, where each cycle exhibits four distinct regimes: (1) a non-normal growth in the $z_1-z_2$ plane, (2) exponential decay in the $z_1-z_2$ plane  to the origin,  (3) an exponential  growth in the $z_3$ direction, and (4) an exponential decay in the $z_3$ direction. In Figure \ref{fig_low}a we present the  trajectory of the dynamical system colored according to the variable $z_3$ in the 3D phase space. The four regimes as described above can be observed. We also present the projection of the vector field in the $z_3=0$ plane, where the non-normal structure can be seen as well. On the other hand, the singularity at $z_1^2+z_2^2=0$ induces a severe exponential growth when the state approaches this region. This configuration allows for the repeated occurrence of non-normal and exponential instabilities.

We note that  due to the exponential instability close to the origin the system undergoes chaotic transitions between positive and negative values of $z_3$. In Figure \ref{fig_low}b we present a single cycle together with a single OTD mode (red vector) shown for different time instants, while the time series for the state variables for one cycle is shown in Figure \ref{fig_low}c. The instantaneous growth rate $\sigma$ corresponding to the computed OTD mode is shown in Figure \ref{fig_low}d. We can clearly observe that the OTD mode initially captures the severe exponential growth and subsequently captures the non-normal growth. On the other hand the eigenvalues of the full linearized operator can only capture the exponential growth, even in regimes where it is not relevant, while they completely miss the non-normal growth.

\subsection{Vertical Jet in Crossflow }
The  \emph{jet in crossflow} is an important problem in fluid mechanics with  a wide range of applications from film cooling of gas turbines,  fuel injection in combustion chambers, to  pollutant  dispersal from chimneys. The interplay of the jet and crossflow creates  coupled vortical structures   whose interactions are highly unsteady and  three dimensional, often leading to turbulence and resulting in a high dimensional dynamical system \cite{HessamPhD}.  The stability of the jet in crossflow has been recently  studied in \cite{bagheri2009global}, where an unstable base flow is computed by forcing the Navier-Stokes equation  to an unstable steady solution  using   the selective frequency damping method \cite{aakervik2006steady}. The Navier-Stokes equation is then linearized around the  base flow and the   global eigenmodes of the linearized Navier-Stokes equation are then computed.

In this section, we compute  the OTD  modes for the vertical jet in crossflow in a weakly turbulent regime.  In particular, we follow the short- and long-time evolution 
 of the OTD subspace with the \emph{time-dependent base flow}, which is obtained by performing  Direct Numerical Simulation (DNS) of the incompressible Navier-Stokes equation.

\subsubsection{Problem specification}\label{JICF-PS}
\begin{figure}
\centering
 \includegraphics[width=0.8\textwidth]{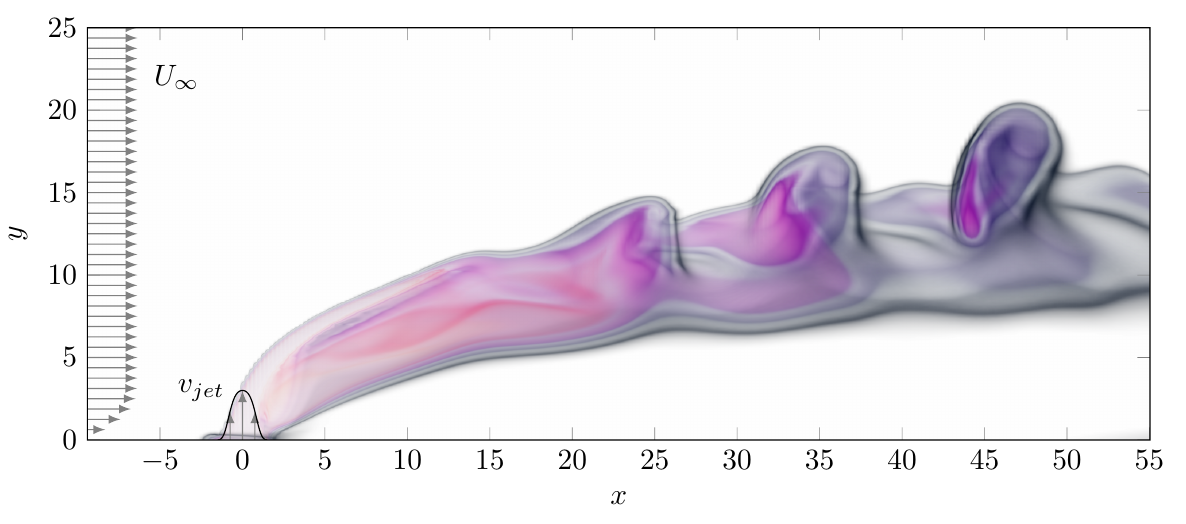}
\caption{Schematic of the vertical jet in crossflow in $x-y$ plane. A snapshot of the base flow is visualized by the volume rendering of the scalar field.} 
\label{JICF-Schem}
\end{figure}

The problem setup is analogous to  several recent studies in the literature \cite{bagheri2009global,HessamMasters}.
 A 2D schematic
of a
vertical jet in crossflow  is shown in Figure \ref{JICF-Schem}, where a vertical
jet is issued into the crossflow.
The characteristic length is the displacement thickness of the crossflow boundary layer.
  The origin of the coordinate system is placed at the center of the jet exit with the jet diameter $D = 3\delta^*$. The computational domain spans from $x=-9.375 \delta^*$ to $x=55 \delta^*$ in the streamwise direction, from
  $y=0$ to $y=50 \delta^*$ in the wall normal direction and from $z=-15 \delta^*$ to $z=15 \delta^*$ in the spanwise direction. 

At the crossflow inlet the Blasius boundary layer profile with the displacement thickness of $\delta^*$ and free-stream velocity of $U_{\infty}$ is specified. The jet velocity profile is given by:
\begin{equation}
v_{jet}(r) = R (1-r^2) \exp( -(r/0.7)^4),
\end{equation}
where $r$ is the normalized distance from the center of the jet:
\begin{equation*}
r = 2/D\sqrt{x^2 +  z^2},
\end{equation*}
and $R=\frac{V_{j}}{U_{\infty}}$ is  the ratio of \emph{peak} jet velocity to the crossflow velocity.
The Reynolds number, based on the crossflow velocity $U_{\infty}$ and the displacement thickness, is given by 
$Re_{\infty} = U_{\infty}\delta^*/\nu$, while the jet Reynolds number is given by $Re_j = V_j D/\nu$. We use a velocity ratio $R=3$, and a Reynolds number $Re_{\infty}=100$ or equivalently $Re_j = 900$.
 At the top boundary the free-stream velocity $\mathbf{U}_b=\{U_{\infty},0,0\}$ is imposed. In the spanwise direction periodic boundary conditions are used. At the outflow boundary a zero-normal gradient is enforced for velocity components.

\subsubsection{OTD equations for Navier-Stokes}
To compute the time-dependent base flow, denoted by $\mathbf{U}_b := \mathbf{U}_b(\mathbf{x},t),$ we solve the incompressible Navier-Stokes equation given by:
\begin{align}\label{NS}
 \frac{\partial{\mathbf{U}_b}}{\partial{t}}  + (\mathbf{U}_b \cdot \nabla ) \mathbf{U}_b &= - \nabla p_b +\dfrac{1}{Re} \nabla^2 \mathbf{U}_b,
    \\
 \nabla \cdot \mathbf{U}_b &= 0, 
\end{align}
along with the boundary conditions given in subsection   \ref{JICF-PS}.
The evolution equation for the modes is given by: 
\begin{align}\label{Sens_NS}
 \frac{\partial{\mathbf{u}_i}}{\partial{t}} &=  \mathcal{L}_{NS}(\mathbf{u}_i) - \big <\mathbf{u}_j, \mathcal{L}_{NS}(\mathbf{u}_i) \big> \mathbf{u}_j
    \\ 
 \nabla \cdot \mathbf{u}_i &= 0  \nonumber
\end{align}
where $\mathcal{L}_{NS}$ is the linearized Navier-Stokes operator, given by:
\begin{align*}
\mathcal{L}_{NS}(\mathbf{u}_i) =  -(\mathbf{U}_b \cdot \nabla ) \mathbf{u}_i - (\mathbf{u}_i \cdot \nabla ) \mathbf{U}_b    +\dfrac{1}{Re} \nabla^2 \mathbf{u}_i - \nabla p_i.
\end{align*}
\begin{figure}
\centering
   \includegraphics[width=0.8\textwidth]{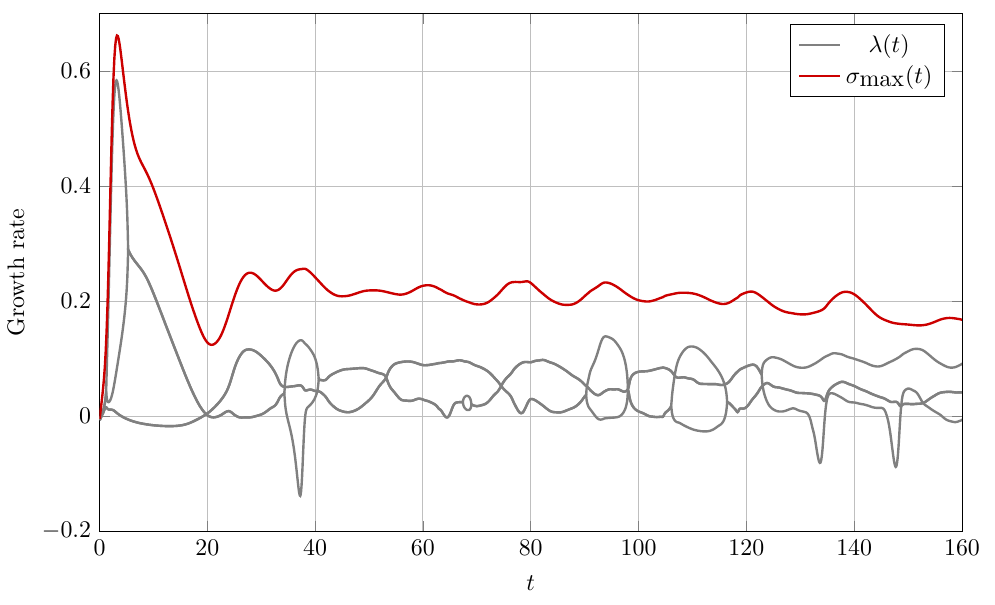}
\caption{Instantaneous real eigenvalues   and the numerical abscissa  of the reduced operator (using the OTD modes), for the
jet in crossflow   at $Re_j=900$.}
 \label{NS-Eig}
\end{figure}A zero boundary  condition for $\mathbf{u}_i$, $i=1, \dots, r$ is enforced at the inflow, wall, jet exit and the top boundaries.
Periodic boundary conditions is used in the spanwise direction and at the outflow, while zero-normal gradient  is imposed on velocity components. In the above evolution equation the choice of the inner product is the second $L_2$ norm in the complex space:
\begin{align}
\big < \mathbf{u} , \mathbf{v}  \big >  =  \int_{\Omega} u_x v_x  + u_y v_y + u_z v_z, 
\end{align}
where $\mathbf{u}= (u_x,u_y,u_z)$ and $\mathbf{v}= (v_x,v_y,v_z)$ are velocity vector fields. The reduced linear operator is therefore obtained from:
\begin{align}
L_{r_{ij}}(t) = \big < \mathbf{u}_i , \mathcal{L}_{NS}(\mathbf{u}_j)  \big >  ,\quad i,j=1, \dots, r. 
\end{align}

\subsubsection{Initial conditions}
The initial condition for the modes is obtained by an orthonormalized space spanned by $\big \{ \mathbf{u}_i(\mathbf{x}) \big \}_{i=1}^r$ with 
$\mathbf{u}_i(\mathbf{x}) = (\partial \psi_i/\partial y , -\partial \psi_i/\partial x, 0)$, where  the two-dimensional streamfunctions $\psi_i$ are chosen as:
\begin{equation}
\psi_i(x,y) = \sin(2\pi f_{x_i} x) \sin(2\pi f_{y_i} y) I(y),
\end{equation}
where $f_{x_i}$ and $f_{y_i}$ are the wavenumbers and $\mathcal{I}(y)$  is a smooth indicator function, localizing the modes in the main body of the jet, i.e. between $y_s=1.0$ and $y_e=6.0$.  More specifically, the indicator function is given by:
\begin{equation}
I(y) = \tanh((y-y_s)/\delta)) - \tanh((y-y_e)/\delta)),
\end{equation}
with $\delta=0.5$. For the calculations that follow we choose a four-dimensional OTD basis, \emph{i.e.} $r=4$.

\subsubsection{Visualization }
For the visualization of the base flow we solve  a passive scalar field
$\theta$ that is governed by the advection-diffusion equation given by:
\begin{equation*}\label{scalar}
 \frac{\partial \theta}{\partial{t}}  + (\mathbf{U}_b \cdot \nabla
) \theta = \frac{1}{Re Sc} \nabla^2 \theta,
\end{equation*}
where $Sc$ is the Schmidt and is chosen to be $Sc=1$. The passive scalar
is set to be $\theta=1$ at the crossflow inlet,  $\theta=0$ at the jet inlet,
periodic condition at spanwise boundaries and zero Neumann  condition on
all other boundaries. As such the jet body region is roughly determined by: 
\begin{equation*}
\mbox{jet body}=\{\mathbf{x}, \mbox{such that} \ 0\leq \theta(\mathbf{x},t)<1\}.
\end{equation*}   
Moreover, by volume-rendering only selected levels of $\theta$, the shear layer and vortical structures can be revealed. In Figure \ref{JICF-Schem}, the volume rendering of $\theta$  exposes the upper and lower shear layers above  the jet exit, and also vortical structures further downstream.
 For visualizing the OTD modes, the iso-surface of the magnitude of velocity of the ranked OTD modes $ \mathbf{U}_i$, colored by the scalar field, is shown.
\subsubsection{Numerical algorithm}
We use a spectral/hp element method to perform Direct Numerical Simulation (DNS)   of the full Navier-Stokes equation and the evolution equation for the OTD basis. The details of the spectral/hp
 element solver \emph{(Nektar)} can be found in  \cite{karniadakis2005spectral}.  We use an unstructured hexahedral mesh with 99792 elements with spectral polynomial of order four. For time integration we use the splitting scheme with first order explicit Euler method with time increments of $\Delta t = 4\times 10^{-3}$. The Navier-Stokes equations are first advanced for 200 time units, by which time the nonlinear dynamical system 
 has reached a statistical steady state.  Due to the inherent similarities of the evolution equations of the OTD basis to the nonlinear Navier-Stokes equation,  the computational cost of solving a system of $r$ OTD modes is roughly equal to $r$ times of a single DNS run. Since the  base is also solved along with the  OTD modes, the total computational  cost  is   $(r+1)$ times of a single DNS run.

\subsubsection{Non-normality  and transient growth}
In Figure \ref{NS-Eig}, the instantaneous  real eigenvalues and the numerical abscissa  of the reduced operator are shown. The large disparity between the numerical abscissa and the largest real eigenvalue of the reduced operator exposes a large degree of non-normality in the reduced operator  $L_r$. The subspace experiences   
significant non-normal growth initially for $0 < t < 20$. This observation is qualitatively in accordance with the linear stability analysis of the jet in crossflow in reference \cite{bagheri2009global}. We refer the reader to Figure 3(c) in that article, in which the initial growth rate of the perturbation is much larger than its asymptotic behavior.

\begin{figure}
\centering
\includegraphics[width=1.0\textwidth]{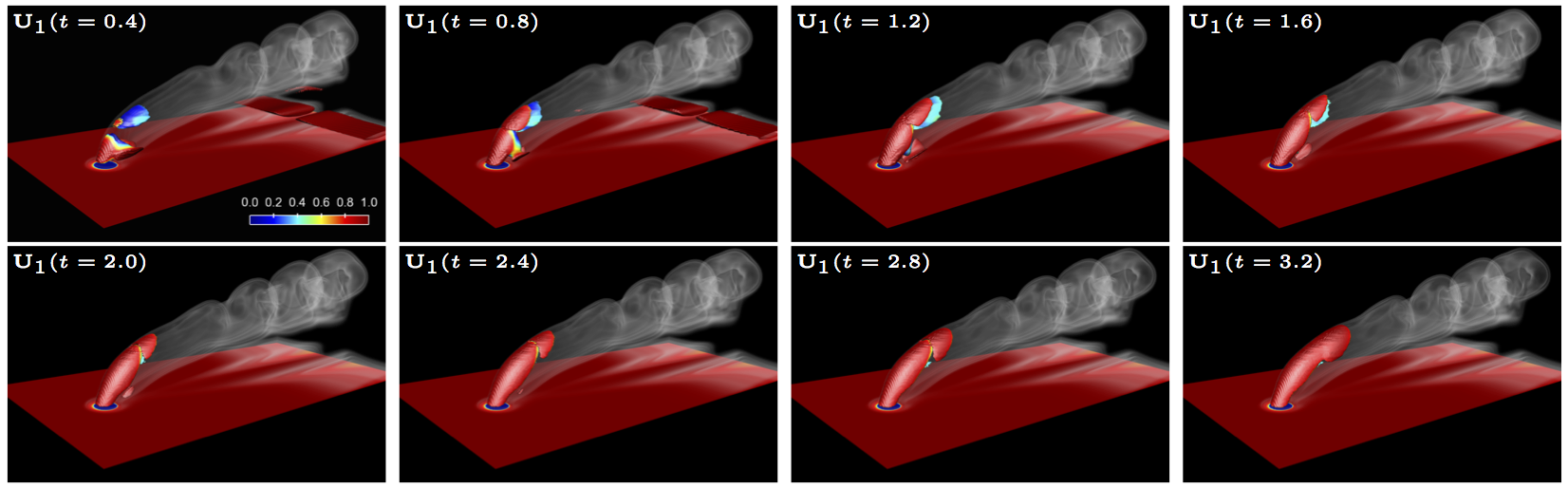}
\caption{ Initial evolution of the first mode  and the trajectory of the Navier-Stokes equations, starting from $t=0.4$ with the  time advancement  of $\Delta t = 0.4$ time units.  The mode is visualized by the iso-surface of the velocity magnitude equal to $0.03$. The time-dependent base flow (DNS) is visualized by smoke volume rendering of a scalar field.}
 \label{JICF-Initial-Mode=1}
\end{figure}

 The snapshots of the initial evolution of the first mode, \emph{i.e.} the most unstable mode, are shown in Figure \ref{JICF-Initial-Mode=1}.  The mode is visualized by the iso-surface of the velocity magnitude equal to $0.03$. At $t=0.4$ the mode clearly captures both  lower and upper shear layers. As time advances, the presence of the upper shear layer becomes more pronounced. This is evident in the snapshots in the second row in Figure \ref{JICF-Initial-Mode=1}.
 It should be reminded that the norm of each mode remains one for all times and as a result the mode quickly vanishes outside the jet body, where the magnitude of $\mathbf{U}_1 (\mathbf{x},t)$
 is significantly smaller relative to the values in  the shear layer regions.  

\subsubsection{Long-time behavior}
      As time progresses, the subspace exhibits \emph{bursts} of growth and sudden excursions into stable directions, as it can be seen in Figure  \ref{NS-Eig}.
At each time instant at least one or more unstable directions can be observed. These unstable directions appear either as single (real eigenvalues) or in pairs (complex conjugate eigenvalues). The unstable 
directions represent \emph{persistent instabilities} that are the hallmark of shear flows.

\begin{figure}[!t]
\centering
\includegraphics[width=1.0\textwidth]{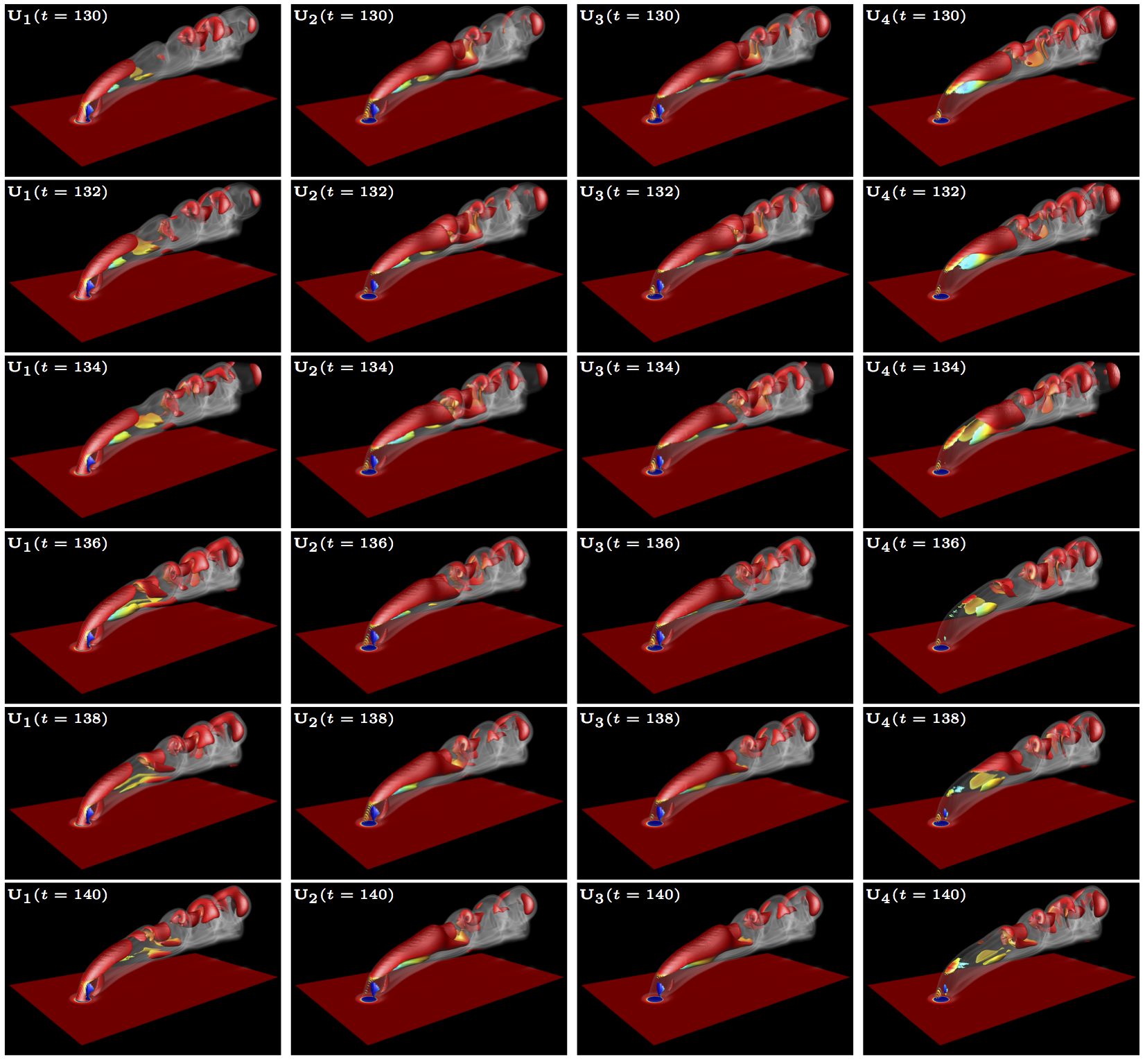}
\caption{Snapshots of the OTD modes $\mathbf{U}_i(\mathbf{x},t)$ and the trajectory of the Navier-Stokes equations. Each row shows all four modes at a given time, with the first row taken at $t=130$. Time advancement  from one row to the next is $\Delta t = 2$ time units.  The modes are visualized by the iso-surface of the velocity magnitude equal to $0.02$. The time-dependent base flow (DNS) is visualized by smoke volume rendering of a scalar field.   }
\label{JICF-Modes}
\end{figure}

In Figure \ref{JICF-Modes} the snapshots of all four modes along with the smoke volume rendering of the Navier-Stokes equation are shown. The modes are visualized
by iso-surfaces of the velocity magnitude (equal to $0.02$), and colored by the scalar field $\theta$.  Each column tracks one mode at different time instants, starting from
the top row at $t=120$, with the time advancement of $\Delta t = 2$ to the next row. The modes are sorted from the most unstable directions (Mode 1) to the most stable directions
 (Mode 4). The first mode captures the \emph{vortex sheet} in the upper shear layer of the jet. This reaffirms the strong evidence that the jet upper shear layer is unstable, leading
 to the vortex roll up further downstream \cite{bagheri2009global}. The second and the third modes show strong presence in both the upper and lower shear 
 layers, while the fourth mode captures the dominant vortical structures downstream. 

The shear layer, spanned by the OTD subspace, is a critical dynamical feature since it is associated with the ``birth place'' of the instability. The strong presence of the upper and lower shear layers in the large times, exposes the important role of non-normality even in the asymptotic dynamics of this flow.
Moreover the upper shear layer
  remains almost steady and as such it has negligible turbulent kinetic energy. Therefore in energy-based reduction techniques such as Proper Orthogonal Decomposition (POD), 
  the shear layer appears strongly  in the time-averaged fields, and only weakly with modes with which unstable directions are associated. A more
comprehensive analysis of the  origin of the modes and their connection with
coherent structures is currently in progress. 

\section{Conclusions}
We have introduced a minimization formulation for the extraction of a finite-dimensional, time-dependent, orthonormal basis, which captures directions of the  phase space associated with transient instabilities. The central idea is to built a set of optimally-time-dependent (OTD) modes with rate of change that optimally spans the vector field of the full dynamical system, in the neighborhood of its current state. We demonstrated how the formulated minimization principle can be utilized to produce evolution equations for these time-dependent modes. These equations require a trajectory of the system, as well as, the linearized operator and their solution gives  a time-dependent, orthonormal basis which spans the current directions (i.e. for the current state of the system)  associated with maximum growth. For the special case of equilibrium states we have shown that these modes rapidly converge to the most unstable directions of the system.

We have demonstrated the capability of the approach on capturing instabilities caused by linear dynamics such as non-normal effects as well as nonlinear exchanges of energy between modes. In particular we have illustrated the computation of the OTD modes in order to capture energy growths/exchanges occurring in: i) linear systems including the advection-diffusion operator in a strongly non-normal regime as well as the Orr-Sommerfeld/Squire operator, and ii) nonlinear systems including a low-dimensional system with both non-normal and exponential growth regimes, and the vertical jet in crossflow in an unstable regime. For the linear systems we demonstrated that the time-dependent subspace captures the strongly transient non-normal energy growth (in the short-time regime), while for longer times the modes capture the expected asymptotic behavior of the dynamics. For the low-dimensional nonlinear system we demonstrated how the subspace captures the most unstable directions of the dynamics, associated with exponential or non-normal growth, while for the fluid flow example we also explored the connection between the shear flow, non-normal growth and persistent instabilities.

The proposed approach paves the way for i) the formulation of efficient, reduced-order filtering and prediction schemes for a variety of infinite dimensional problems involving strongly transient features, such as rare events, and ii) the formulation of low-energy control algorithms that will be able to suppress  the instability in a very early stage by applying reduced-order control methods the moment that the instability has begun to emerge. The proposed framework should also be important for the fundamental understanding of the dynamical processes behind transient features, through the computation of finite-time Lyapunov exponents (a task that is not feasible in an infinite dimensional setting) and the analysis of the associated energy transfers.
 
\subsubsection*{Acknowledgments} The authors would like to thank Prof. G. Karniadakis , Dr. M. Farazmand, and Mr. S. Mowlavi for discussions that led to a number of improvements. TPS has been supported through the Army Research Office Young Investigator Award  66710-EG-YIP, the Office of Naval Research grant ONR N00014-14-1-0520, and the DARPA grant HR0011-14-1-0060. HB has been partially supported as a post-doc by the first and third grants. The authors  gratefully acknowledge the allocated computer time on the Stampede supercomputers, awarded by the XSEDE program, allocation number TG-ECS140006. 
\vskip6pt

\bibliographystyle{plain}
\bibliography{biblio,library}

\begin{thebibliography}{10}

\bibitem{chebfun}
Driscoll~T. A., Hale N., and Trefethen~L. N.
\newblock Chebfun guide, pafnuty publications.
\newblock Technical report, Oxford, 2014.

\bibitem{aakervik2006steady}
E.~{\AA}kervik, L.~Brandt, D.S. Henningson, J.~H{\oe}pffner, O.~Marxen, and
  P.~Schlatter.
\newblock Steady solutions of the navier-stokes equations by selective
  frequency damping.
\newblock {\em Physics of Fluids}, 18:068102, 2006.

\bibitem{akhm13}
N.~Akhmediev, J.~M. Dudley, D.~R. Solli, and S.~K. Turitsyn.
\newblock {Recent progress in investigating optical rogue waves}.
\newblock {\em Journal of Optics}, 15(6):60201, 2013.

\bibitem{arec11}
F.~T. Arecchi, U.~Bortolozzo, A.~Montina, and S.~Residori.
\newblock {Granularity and inhomogeneity are the joint generators of optical
  rogue waves}.
\newblock {\em Physical Review Letters}, 106(15):2--5, 2011.

\bibitem{arnold_l}
L.~Arnold, I.~Chueshov, and G.~Ochs.
\newblock {Stability and capsizing of ships in random sea - A survey}.
\newblock {\em Nonlinear Dynamics}, 36:135--179, 2004.

\bibitem{Avila08072011}
K.~Avila, D.~Moxey, Alberto D.~L., M.~Avila, D.~Barkley, and B.~Hof.
\newblock The onset of turbulence in pipe flow.
\newblock {\em Science}, 333(6039):192--196, 2011.

\bibitem{HessamPhD}
H.~Babaee.
\newblock {\em Analysis and optimization of film cooling effectiveness}.
\newblock PhD thesis, Louisiana State University, August 2013.

\bibitem{HessamMasters}
H.~Babaee.
\newblock Uncertainty quantification of film cooling effectiveness in gas
  turbines.
\newblock Master's thesis, Louisiana State University, 2013.

\bibitem{bagheri2009global}
S.~Bagheri, P.~Schlatter, P.~J. Schmid, and D.~S. Henningson.
\newblock Global stability of a jet in crossflow.
\newblock {\em Journal of Fluid Mechanics}, 624:33--44, 2009.

\bibitem{Boberg-and-Brosa}
L.~B{\"o}berg and U.~Br{\"o}sa.
\newblock Onset of turbulence in a pipe.
\newblock {\em Zeitschrift f{\"u}r Naturforschung A.}, 43a(697), 1988.

\bibitem{Bourlioux2002}
A.~Bourlioux and A.~J. Majda.
\newblock {Elementary models with probability distribution function
  intermittency for passive scalars with a mean gradient}.
\newblock {\em Physics of Fluids}, 14(2):881--897, 2002.

\bibitem{Bourlioux2006}
A.~Bourlioux, A.~J. Majda, and O.~Volkov.
\newblock {Conditional statistics for a passive scalar with a mean gradient and
  intermittency}.
\newblock {\em Physics of Fluids}, 18(10):1--10, 2006.

\bibitem{branic_majda}
M.~Branicki and A.~J. Majda.
\newblock {Quantifying uncertainty for predictions with model error in
  non-Gaussian systems with intermittency}.
\newblock {\em Nonlinearity}, 25:2543, 2012.

\bibitem{Farrel_92}
K.~M. Butler and B.~F. Farrell.
\newblock Three dimensional optimal perturbations in viscous shear flow.
\newblock {\em Physics of Fluids A}, 4(8):1637--1650, 1992.

\bibitem{chandler13}
G.~J. Chandler and R.~R. Kerswell.
\newblock {Invariant recurrent solutions embedded in a turbulent
  two-dimensional Kolmogorov flow}.
\newblock {\em J. Fluid Mech.}, 722:554--595, 2013.

\bibitem{Hou13a}
M.~Cheng, T.~Hou, and Z.~Zhang.
\newblock {A dynamically bi-orthogonal method for time-dependent stochastic
  PDEs I: Derivation and algortihms}.
\newblock {\em Journal of Computational Physics}, 242:843--868, 2013.

\bibitem{Cornelius2013}
S.~P. Cornelius, Wi.~L. Kath, and A.~E. Motter.
\newblock {Realistic control of network dynamics.}
\newblock {\em Nature communications}, 4:1942, 2013.

\bibitem{cousins_sapsis}
W.~Cousins and T.~P. Sapsis.
\newblock {Quantification and prediction of extreme events in a one-dimensional
  nonlinear dispersive wave model}.
\newblock {\em Physica D}, 280:48--58, 2014.

\bibitem{cousinsSapsis2015_JFM}
W.~Cousins and T.~P. Sapsis.
\newblock {Reduced order prediction of rare events in unidirectional nonlinear
  water waves}.
\newblock {\em Submitted to Journal of Fluid Mechanics}, 2015.

\bibitem{cousinsSapsis2015_PRE}
W.~Cousins and T.~P. Sapsis.
\newblock {The unsteady evolution of localized unidirectional deep water wave
  groups}.
\newblock {\em Physical Review E}, 91:063204, 2015.

\bibitem{Egolf:2000aa}
D.~A. Egolf, I.~V. Melnikov, W.~Pesch, and Robert~E. E.
\newblock Mechanisms of extensive spatiotemporal chaos in rayleigh-benard
  convection.
\newblock {\em Nature}, 404(6779):733--736, 04 2000.

\bibitem{farrell_2}
B.~F. Farrell and P.~J. Ioannou.
\newblock {Generalized stability theory Part II: non-autonomous operators}.
\newblock {\em J. Atmos. Sci.}, 53:2041--2053, 1996.

\bibitem{grab01}
W.~W. Grabowski.
\newblock {Coupling Cloud Processes with the Large-Scale Dynamics Using the
  Cloud-Resolving Convection Parameterization (CRCP)}.
\newblock {\em Journal of the Atmospheric Sciences}, 58(9):978--997, May 2001.

\bibitem{grab99}
W.~W. Grabowski and P.~K. Smolarkiewicz.
\newblock {CRCP: a Cloud Resolving Convection Parameterization for modeling the
  tropical convecting atmosphere}.
\newblock {\em Physica D: Nonlinear Phenomena}, 133(1-4):171--178, 1999.

\bibitem{Haller2010}
G.~Haller and T.~Sapsis.
\newblock {Localized Instability and Attraction along Invariant Manifolds}.
\newblock {\em SIAM Journal on Applied Dynamical Systems}, 9(2):611--633, 2010.

\bibitem{hamilton95}
J.~M. Hamilton, J.~Kim, and F.~Waleffe.
\newblock {Regeneration mechanisms of near-wall turbulence structures}.
\newblock {\em J. Fluid Mech.}, 287:243, 1995.

\bibitem{FLM:339375}
D.~S. Henningson, A.~Lundbladh, and A.~V. Johansson.
\newblock A mechanism for bypass transition from localized disturbances in
  wall-bounded shear flows.
\newblock {\em Journal of Fluid Mechanics}, 250:169--207, 5 1993.

\bibitem{karniadakis2005spectral}
G.~E. Karniadakis and S.~J. Sherwin.
\newblock {\em Spectral/hp element methods for computational fluid dynamics}.
\newblock Oxford University Press, USA, 2005.

\bibitem{vakakis_kourdis}
P.~D. Kourdis and A.~F. Vakakis.
\newblock {Some results on the dynamics of the linear parametric oscillator
  with general time-varying frequency}.
\newblock {\em Applied Mathematics and Computation}, 183:1235--1248, 2006.

\bibitem{majda2000}
A.~J. Majda.
\newblock {Real world turbulence and modern applied mathematics}.
\newblock In {\em Mathematics: Frontiers and Perspectives, International
  Mathematical Union}, pages 137--151. American Mathematical Society, 2000.

\bibitem{majda11}
A.~J. Majda.
\newblock {Challenges in Climate Science and Contemporary Applied Mathematics}.
\newblock {\em Communications on Pure and Applied Mathematics}, 65:920, 2012.

\bibitem{majda_branicki_DCDS}
A.~J. Majda and M.~Branicki.
\newblock {Lessons in Uncertainty Quantification for Turbulent Dynamical
  Systems}.
\newblock {\em Discrete and Continuous Dynamical Systems}, 32:3133--3221, 2012.

\bibitem{Majda_filter}
A.~J. Majda and J.~Harlim.
\newblock {\em {Filtering Complex Turbulent Systems}}.
\newblock Cambridge University Press, 2012.

\bibitem{mohamad2015}
M.~A. Mohamad and T.~P. Sapsis.
\newblock {Probabilistic description of extreme events in intermittently
  unstable systems excited by correlated stochastic processes}.
\newblock {\em SIAM ASA J. of Uncertainty Quantification}, 3:709--736, 2015.

\bibitem{mohamad2015a}
M.~A. Mohamad and T.~P. Sapsis.
\newblock {Probabilistic response of Mathieu equation excited by correlated
  parametric excitation}.
\newblock {\em Ocean Engineering Journal - Submitted}, 2015.

\bibitem{Moxey04052010}
D.~Moxey and D.~Barkley.
\newblock Distinct large-scale turbulent-laminar states in transitional pipe
  flow.
\newblock {\em Proceedings of the National Academy of Sciences},
  107(18):8091--8096, 2010.

\bibitem{muller}
P.~Muller, C.~Garrett, and A.~Osborne.
\newblock {Rogue Waves}.
\newblock {\em Oceanography}, 18(3):66--75, 2005.

\bibitem{onorato13}
M.~Onorato, S.~Residori, U.~Bortolozzo, A.~Montina, and F.~T. Arecchi.
\newblock {Rogue waves and their generating mechanisms in different physical
  contexts}.
\newblock {\em Physics Reports}, 528(2):47--89, 2013.

\bibitem{orszag1971accurate}
S.A. Orszag.
\newblock Accurate solution of the orr-sommerfeld stability equation.
\newblock {\em J. Fluid Mech}, 50(4):689--703, 1971.

\bibitem{Pastoor2008}
M.~Pastoor, L.~Henning, B.~R. Noack, R.~King, and G.~Tadmor.
\newblock {Feedback shear layer control for bluff body drag reduction}.
\newblock {\em Journal of Fluid Mechanics}, 608:161--196, 2008.

\bibitem{pope97}
S.~B. Pope.
\newblock {Computationally efficient implementation of combustion chemistry
  using 'in situ' adaptive tabulation}.
\newblock {\em Combustion Theory and Modelling}, 1(1):41--63, January 1997.

\bibitem{doi:10.1137/0153002}
S.~C. Reddy, P.~J. Schmid, and D.~S. Henningson.
\newblock Pseudospectra of the orr-sommerfeld operator.
\newblock {\em SIAM Journal on Applied Mathematics}, 53(1):15--47, 1993.

\bibitem{RevModPhys.72.603}
Grossmann S.
\newblock The onset of shear flow turbulence.
\newblock {\em Rev. Mod. Phys.}, 72:603--618, Apr 2000.

\bibitem{sapsis11a}
T.~P. Sapsis.
\newblock {Attractor local dimensionality, nonlinear energy transfers, and
  finite-time instabilities in unstable dynamical systems with applications to
  2D fluid flows}.
\newblock {\em Proceedings of the Royal Society A}, 469(2153):20120550, 2013.

\bibitem{sapsisdijkstra}
T.~P. Sapsis and H.~A. Dijkstra.
\newblock {Interaction of noise and nonlinear dynamics in the double-gyre
  wind-driven ocean circulation}.
\newblock {\em J. Phys. Oceanography}, 43:366--381, 2013.

\bibitem{SapsisLermusiaux09}
T.~P. Sapsis and P.~F.~J. Lermusiaux.
\newblock {Dynamically Orthogonal field equations for continuous stochastic
  dynamical systems}.
\newblock {\em Physica D}, 238:2347--2360, 2009.

\bibitem{SapsisLermusiaux10}
T.~P. Sapsis and P.~F.~J. Lermusiaux.
\newblock {Dynamical criteria for the evolution of the stochastic
  dimensionality in flows with uncertainty}.
\newblock {\em Physica D}, 241:60, 2012.

\bibitem{sapsis_DONS}
T.~P. Sapsis, M.~P. Ueckermann, and P.~F.~J. Lermusiaux.
\newblock {Global analysis of Navier-Stokes and Boussinesq stochastic flows
  using dynamical orthogonality}.
\newblock {\em J. Fluid Mech.}, 2013.

\bibitem{Schmid:2000aa}
P.~J. Schmid.
\newblock Linear stability theory and bypass transition in shear flows.
\newblock {\em Physics of Fluids}, pages 1788--1794, 2000.

\bibitem{schmid2007nonmodal}
P.~J. Schmid.
\newblock Nonmodal stability theory.
\newblock {\em Annu. Rev. Fluid Mech.}, 39:129--162, 2007.

\bibitem{Henningson_Schmidt}
P.~J. Schmid and D.~S. Henningson.
\newblock {\em Stability and Transition Stability in Shear Flows}.
\newblock Springer, 2001.

\bibitem{smooke86}
M.~D. Smooke, R.~E. Mitchell, and D.~E. Keyes.
\newblock {Numerical Solution of Two-Dimensional Axisymmetric Laminar Diffusion
  Flames}.
\newblock {\em Combustion Science and Technology}, 67(4-6):85--122, October
  1986.

\bibitem{Susuki2012}
Y.~Susuki and I.~Mezi\'{c}.
\newblock {Nonlinear koopman modes and a precursor to power system swing
  instabilities}.
\newblock {\em IEEE Transactions on Power Systems}, 27(3):1182--1191, 2012.

\bibitem{Susuki2014}
Y.~Susuki and I.~Mezic.
\newblock {Nonlinear Koopman Modes and Power System Stability Assessment
  without Models}.
\newblock {\em IEEE Transactions on Power Systems}, 29(2):899--907, 2014.

\bibitem{tadmor11}
G.~Tadmor, O.~Lehmann, B.~R. Noack, L.~Cordier, J.~Delville, J.~P. Bonnet, and
  M.~Morzynski.
\newblock {Reduced order models for closed-loop wake control}.
\newblock {\em Philosophical Trans. Royal Soc. A}, 369:1513--1524, 2011.

\bibitem{Tantet2015}
A.~Tantet, F.~R. van~der Burgt, and H.~A. Dijkstra.
\newblock {An early warning indicator for atmospheric blocking events using
  transfer operators.}
\newblock {\em Chaos (Woodbury, N.Y.)}, 25(3):036406, March 2015.

\bibitem{Tong2015}
X.~Tong and A.~J. Majda.
\newblock {Intermittency in Turbulent Diffusion Models with a Mean Gradient}.
\newblock {\em Nonlinearity-Submitted}, 2015.

\bibitem{Tref_Pseudo}
L.~N. Trefethen.
\newblock Pseudospectra of linear operators.
\newblock {\em SIAM Review}, 39(3):383--406, 1997.

\bibitem{Trefethen30071993}
L.~N. Trefethen, A.~E. Trefethen, S.~C. Reddy, and T.~A. Driscoll.
\newblock Hydrodynamic stability without eigenvalues.
\newblock {\em Science}, 261(5121):578--584, 1993.

\end{thebibliography}

\end{document}